\documentclass[12pt]{report}
\usepackage{graphicx, subfigure}
\usepackage{amsmath}
\usepackage{amssymb}
\usepackage{cite}

\title{Evolution of intratumoral phenotypic heterogeneity: the role of trait inheritance}
\author{
        Jill Gallaher and Alexander R.\ A.\ Anderson	 \\
        Department of Mathematical Oncology\\
        H. Lee Moffitt Cancer Center\\
        Tampa, FL 33612
}
\date{\today}

\begin{document}
\maketitle

\section*{Abstract}  
{\small A tumor is a heterogeneous population of cells that competes for limited resources. In the clinic, we typically probe the tumor by biopsy, and then characterize it by the dominant genetic clone. But genotypes are only the first link in the chain of hierarchal events that leads to a specific cell phenotype. The relationship between genotype and phenotype is not simple, and the so-called genotype to phenotype map is poorly understood. Many genotypes can produce the same phenotype, so genetic heterogeneity may not translate directly to phenotypic heterogeneity. We therefore choose to focus on the functional endpoint, the phenotype as defined by a collection of cellular traits (e.g.\ proliferative and migratory ability). Here we will examine how phenotypic heterogeneity evolves in space and time and how the way in which phenotypes are inherited will drive this evolution.

A tumor can be thought of as an ecosystem, which critically means that we cannot just consider it as a collection of mutated cells but more as a complex system of many interacting cellular and microenvironmental elements. At its simplest, a growing tumor with increased proliferation capacity must compete for space as a limited resource. Hypercellularity leads to a contact-inhibited core with a competitive proliferating rim. Evolution and selection occurs, and an individual cell's capacity to survive and propagate is determined by its combination of traits and interaction with the environment. With heterogeneity in phenotypes, the clone that will dominate is not always obvious as there are both local interactions and global pressures. Several combinations of phenotypes can coexist, changing the fitness of the whole. 

To understand some aspects of heterogeneity in a growing tumor we build an off-lattice agent based model consisting of individual cells with assigned trait values for proliferation and migration rates. We represent heterogeneity in these traits with frequency distributions and combinations of traits with density maps. How the distributions change over time is dependent on how traits are passed on to progeny cells, which is our main inquiry. We bypass the translation of genetics to behavior by focussing on the functional end result of inheritance of the phenotype combined with the environmental influence of limited space.}

\section{Introduction}
\normalsize 
Tumors are phenotypically and genotypically heterogeneous. Understanding even the most superficial definitions of a tumor's complexity may seem an overwhelming and daunting task, yet characterizing this heterogeneity has begun to gain popularity in recent years \cite{Merlo10,Marusyk12,Gerlinger12,Sottoriva13,Marusyk13}. The pursuit of embracing the complexity of the myriad of possible interactions between individual parts is gaining ground in computational and mathematical oncology, and we are learning how to represent the tumor as the heterogeneous system that it is. 

With heterogeneity comes a possibility for competition and selection. A growing tumor is limited for space and resources, so not all individual cells will continue propagating over time. The cellular population may adapt for enhanced overall fitness, but selection can also be used to our advantage when it comes to treatment if we can better understand how the interacting parts combine to form the emergent whole.

The actual function of each cell is not just determined by genotype \cite{Kreso13}. The same genotype may yield different phenotypes and the same phenotype may be borne from multiple genotypes \cite{Gerlinger12,Kreso13}. Add to that complexity the epigenetic modulations that affect the cell's response \cite{Rando07}, the microenvironment \cite{Anderson06,Anderson09}, differential protein expression and biochemical noise \cite{Niepel09,Brock09,Spencer09,Ma11} and the problem of mapping genotype to phenotype appears to be impossible \cite{Burga12}. To reduce this inherent complexity and because the cell's function is what selection will act on, we focus on the phenotype.  

Here we will present a method that will characterize the key elements of phenotypic heterogeneity in a growing tumor and simulate situations where the cells that make up this tumor will compete with each other. We investigate, using the traits of proliferation and migration, 1) how trait inheritance matters, 2) how spatial competition affects trait selection, and 3) how constraints on trait combinations affect population growth.

With heterogeneity in the phenotypes of a collection of cells, it matters how a cell passes on these traits. The role of phenotypic inheritance in tumor progression under environmental stresses has been considered before, but remains an open question \cite{Lachmann96,Staudte97,Rando07,Anderson09,Quaranta09,Feinberg10,Sottoriva11,Guerrero12}. In the pioneering work by Anderson and colleagues, the evolution of phenotypic heterogeneity of five different traits was investigated.  They compared a linear progression of mutation and a more random acquisition of traits (representing phenotypic plasticity) in a growing tumor in a range of different microenvironments \cite{Anderson05,Anderson06}. We continue this line of thought by exploring trait inheritance in a much simpler model with only two traits, namely proliferation and migration, under the stress of limited space. 

Space is the only limited resource we consider here. We assume that with hypercellularity the cells are subject to contact inhibition which induces quiescence. We look at two different configurations that lead to different degrees of competition: a dispersion of cells and a cluster of cells. The dispersion of cells begin spread out and form their own colonies with plenty of space. This may be analogous to a cell suspension in vitro or an in vivo situation where competition is less pronounced (e.g.\ circulating tumor cells). The cluster of cells start tightly packed together so that as it grows only the cells on the rim of the mass will be in the proliferating state. Contact inhibition will play a bigger role in providing a selective pressure for evolution of the population in the latter configuration. In a model by Lee et al, the growth of populations in similar spatial configurations was investigated and the importance of contact inhibition was highlighted \cite{Lee95}, but this model did not consider heterogeneity or variations in the mode of inheritance, which will be our primary focus here. 

With limited energy and resources, the combinations of traits that allow cell viability will be bounded. Much has been studied on the idea of the proliferation-migration dichotomy (i.e.\ go-or-grow) \cite{Giese03,Mansury06,Hatzikirou12}, where there is a trade-off between proliferating and migrating. This is a simplistic view of a complicated set of possibilities, but there are some combinations of traits that a cell will never achieve. Many traits are coupled in such a way that an effect on one is sure to impact another. As more traits are considered, the complexity increases dramatically. 
   
This study also provokes speculation for how treatment might be affected by the heterogeneity within a tumor and the inheritance mode of the cells. To keep things as simple as possible we will not include cell death in this model, but the implications remain that wiping out a particular cell type will leave room and resources for other non-targeted cell types to take over \cite{Greaves12}. The selection of phenotypes in a growing population with environmental influences is an important concept that should not be ignored. Most standard of care anti-cancer therapies act on phenotypic traits (e.g.\ anti-mitotics, anti-proliferatives, radiation, chemotherapies, etc.). So we are interested in seeing how the underlying lineage of diversity comes about in the first place. 

This simple model provides a brief look at how inheritance of proliferation rates and migration rates affects the overall heterogeneity and fitness of a growing population of cells. We build an off-lattice agent based model to inspect how the many interacting cells compete for space as they pass on their traits in various ways. In the first simulations, we vary each trait individually to get an idea of how they alone influence the collective behavior of the population. The subsequent simulations allow combinations of traits with different constraints imposed on the phenotype space. After discussing  how the heterogeneity of a population is affected by inheritance, we analyze how the different schemes affect the fitness of the population.

\section{The model}
To capture the competition between individuals with heterogeneity in proliferation and migration rates, we implement an agent based model.  Each cell is given a set of traits initially, but we also need to keep a memory of the previous traits that are important for some inheritance modes. Each cell at any time is defined by a set of parameters given by:

\begin{equation}
[p,m]=[(\rho,\rho_0),(\nu,\nu_0,p_T,\theta)].
\end{equation}

\noindent The phenotype of the cell is represented by a combination of trait states; the proliferation $p$ state and the migration state $m$. The proliferation parameter indicates how quickly a cell moves through the cell cycle in terms of an intermitotic time (IMT), and the migration parameters indicate how fast a cell is moving (speed), how long it moves in the same direction (persistence time), and in which direction it is moving. For a given cell, we define the proliferation parameters, $\rho$ and $\rho_0$, as the current and previous proliferation rates, respectively. The current and previous migration speeds are $\nu$ and $\nu_0$, respectively, $p_T$ is the persistence time, and $\theta$ is the angular direction in which the cell is moving. 

\begin{figure}[ht!]
   \centering
            \includegraphics[width=\textwidth]{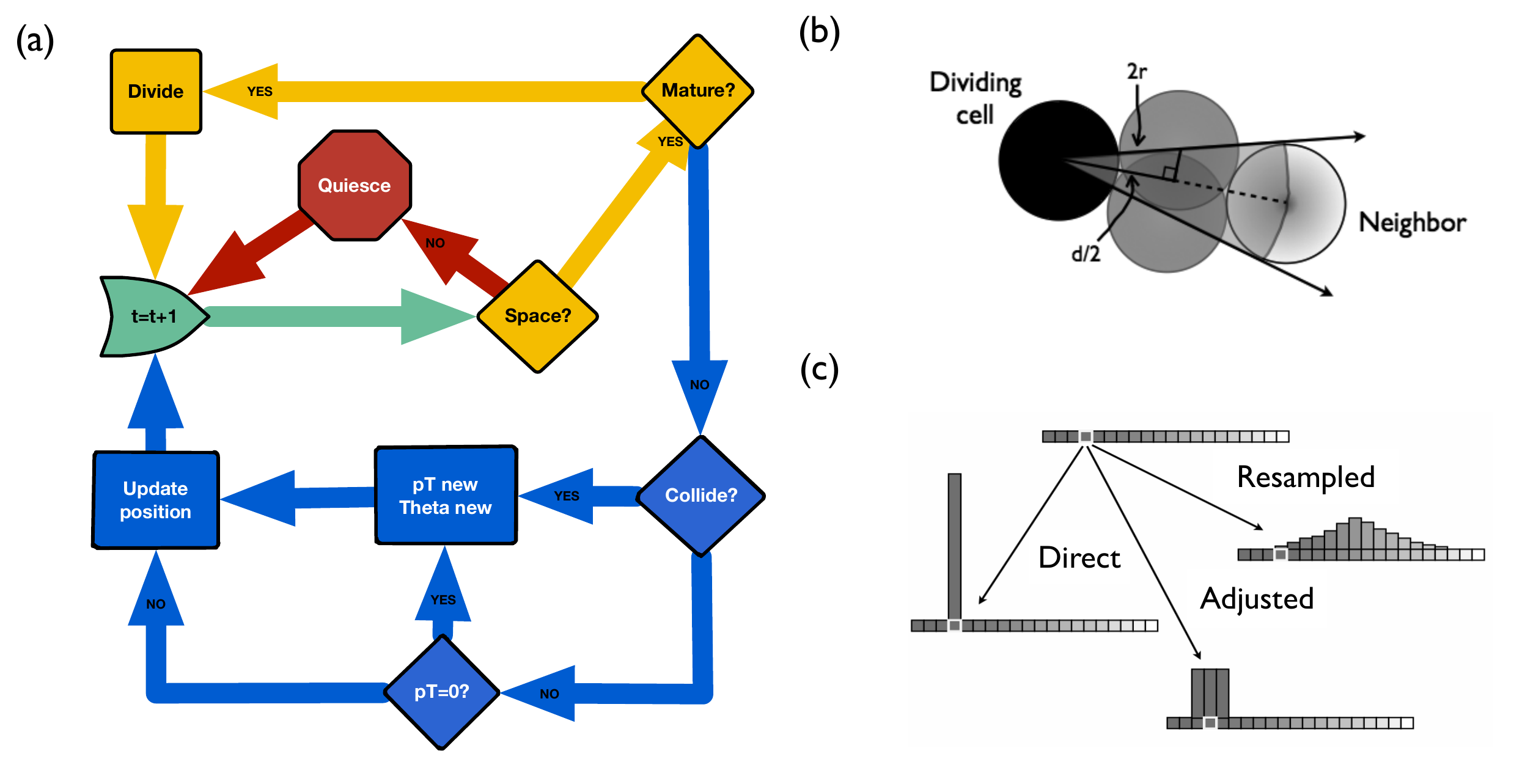} 
            \caption{%
        Model details: a) Flow diagram describing decision making of each cell at each time frame. The time frame starts at the green arrow where it either goes through division (yellow loop), goes into quiescence (red loop), or migrates (blue loop). b) The geometry of angle exclusion that leads to quiescence from contact inhibition from a dividing cell (black) and its neighbor (light gray). The mid-gray ``phantom" cells show the closest possible position of a new cell on each side of the overlap so that there is a block of excluded angles. c) The different modes of inheritance (direct, adjusted, and resampled) are shown with the probability of choosing the trait value of the daughter cells given the parent's trait value (white box). 
     }%
     \label{flowchart}
\end{figure}

For the simulation each cell is given a random starting age in the cell cycle, and for each time step, it will follow the decision tree presented in Fig.\ \ref{flowchart}a. The simulation starts at the green shape marked ``t=t+1", which signifies the start of a new time step. At each time frame, the cell either divides (yellow loop) if it is ready, goes quiescent if it has too many neighbors (red loop), or migrates (blue loop). If it has enough space, the cell will spend most of its time in the migration loop whilst aging and only go through the act of division during a single time step. In the quiescent state, the cell discontinues aging and moving.  

All simulations follow the same rules for proliferation and migration, but vary in the inheritance mode, the spatial layout, and the combinations of traits permitted. The cell moves by an angle of direction and a speed instead of residing on an orthogonal lattice; the angular lattice is resolved to $\Delta \theta = 1$ degree, and the updating occurs with a time step of 1 minute. All cells are 20 microns in diameter.

\subsection{Proliferation and quiescence}
The cells are each given an intermitotic time (IMT), i.e.\ the time that it takes to go through division.  At each time point each cell checks for close neighbors by searching its surrounding space for other cells positioned at a distance ($d$) less than four cell radii ($r$) away (a position close enough for possible overlap with a newly formed daughter cell). For each of these close neighbors, a block of angles $\Delta \theta = \arccos(d/4r)$ to each side of the bisector of the cell is excluded from the bank of allowable angles (c.f.\ Fig.\ \ref{flowchart}b), removing the possibility of overlap upon division. If at any point a cell runs out of available angles, it will go into quiescence, which means here that it stops migrating and stops progressing through the cell cycle. If the quiescent cell has room at a subsequent time to proliferate as neighbors move out of the way, it will pick up where it left off in the cell cycle with its previous migration parameters. The domain boundaries are handled just like neighboring cells, by simply taking up space and excluding available angles.

When it is time to divide, if there is space, one daughter cell will take up the space previously occupied by the parent, and the second will divide into a neighboring space. The new angle is chosen randomly from all available angles. Given this new angle $\theta$, the newly positioned cell will lie a cell diameter away in that direction. During the division time step, the cell will not move. Until then, if not quiescent, the cell migrates.

\subsection{Migration \& cell collisions}
Because migration plays such a large role here, we set up an off-lattice model to capture the more subtle nuances in movement. We assume that the cells follow a persistent random walk \cite{Lee95,Codling08,Quaranta09,Jeon10}. In this case, the cell has a ``preferred" or most likely amount of time it will spend continuing in the same direction. We assume each cell has a predetermined persistence time, $p_T$, that is randomly drawn from a distribution (normally distributed around 80 minutes with a standard deviation of 10 minutes).  If the cell hasn't divided or collided with another cell or the boundary before the persistence time is reached, it will assume a new direction at random and obtain a new persistence time from the distribution. 

At each time frame, we loop through all cells currently in the migratory state and identify if any overlap has occurred with another cell. The time step is sufficiently small that the fastest moving cells will detect an overlap just under $0.5 \mu$m. If two cells are found to overlap, then a collision response begins at the time of first contact within the time frame, then the cells move for the remaining time in the frame  as usual with the new trajectory given by the angle of collision. Collisions of one cell with another follow a regular elastic collision response, i.e.\ the new angle is reflected along the normal. The cells collide with the domain boundaries in the same way. After a collision, a new persistence time is obtained from the persistence distribution for each cell. If the cell divides, each daughter cell obtains a new persistence from the distribution, but they move apart in opposite directions along their normal instead. In the absence of a collision or division, the cell simply continues in the same direction at its given speed. 

Variation in the distribution of persistence times and turning angles do result in differences in population scale dynamics, but in order to keep this manuscript focussed they will not be investigated here. We assume that these parameters are less determined by the cell itself and more by the microenvironment. A very directional type of environment (e.g.\ muscle fibers or blood vessels) might cause a cell to move with long persistences and small changes in turning angles, whereas a more tortuous environment (e.g.\ gray matter in the brain) might subject a cell to smaller persistences and a more random distribution of turning angles. We therefore don't consider these to be appropriate as heritable traits, like the proliferation rate and the migration speed.  

\subsection{Inheritance Schemes}
In this study, we primarily investigate the role of trait inheritance. Three schemes that represent different modes of adaptability will be presented. We refer to the inheritance schemes as: direct, adjusted, and resampled. Figure \ref{flowchart}c relates these various modes of inheritance as a probability that a trait will assume a new value either the same as, close to, or completely different from the parent's current value. 

With direct inheritance, daughters inherit exactly the same trait value as the parent. We therefore define the probability $P$ of the daughter cell inheriting the parent's trait $\tau_p$ as simply:

\begin{equation}
P(\tau_d) = 1 \text{\space if \space} \tau_d=\tau_p\text{.}
\end{equation}

With adjusted inheritance, the trait value $\tau_d$ of the daughter cells may drift slightly from the parent's trait in either direction by a small value $\epsilon$, where $\epsilon$ is the size of one bin in the discretized distribution of allowed values. So the probability that the daughter cell inherits a trait is:
 
\begin{equation}
P(\tau_d)= 
\begin{cases} 1/3& \text{if } \tau_d=\tau_p-\epsilon \text{,}\\
1/3 &\text{if } \tau_d=\tau_p \text{,}\\
1/3& \text{if } \tau_d=\tau_p+\epsilon \text{.}\\
\end{cases}
\end{equation}

\noindent The values are confined to a range, so surpassing an upper or lower bound for the adjusted inheritance, leads to the retention of the previous trait at that bound. 

With resampled inheritance, the new trait values for the daughters will have no memory of the parent's trait, but instead will be randomly chosen from a weighted normal distribution: 
   
\begin{equation}
P(\tau_d) = f(\tau_d) = \frac{1}{\sqrt{2\pi \sigma^2}}e^{(\tau_d-\mu)^2/2\sigma^2}.
\end{equation}

\noindent where $\mu$ is the mean and $\sigma$ is the standard deviation. In all of the following simulations, the resampling distribution is the same as the initial distribution of trait values. Using these modes of inheritance, we will investigate their effects on the heterogeneity of a growing cell population over space and time. 

\section{Results}
\begin{figure}[ht!]
   \centering
            \includegraphics[width=\textwidth]{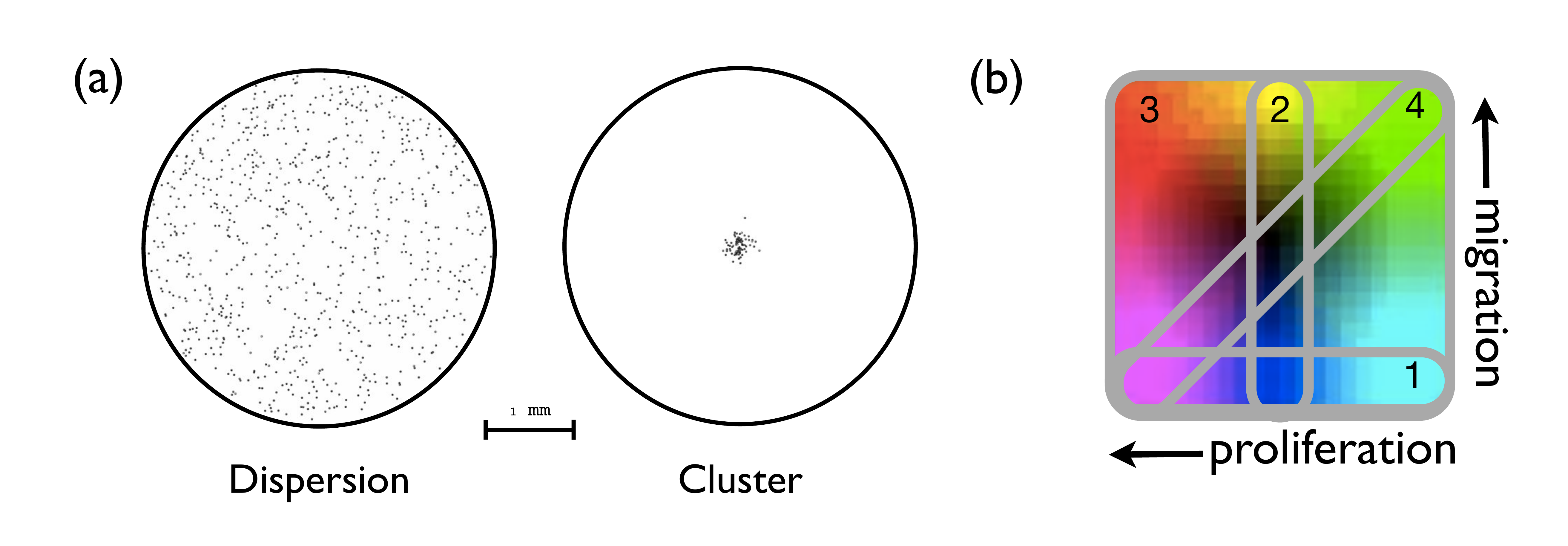} 
            \caption{%
        Model details: a) The two initial spatial configurations: dispersion and cluster. b) The different constraints on the phenotype space labelled in order of presentation (1: proliferation only, 2: migration only, 3: no constraints, 4: go-or-grow)
     }%
     \label{constraints}
\end{figure}

The three previously introduced inheritance schemes are examined using two different spatial configurations. We refer to these as a cellular dispersion and a cellular cluster. With the cellular dispersion, 800 cells are placed randomly within a circular area 4 mm in diameter, which could represent either an in vitro configuration or a disseminated cancer with many seeds spatially spread out. The simulation runs until reaching a population of 25,000 cells, which is sufficient to still discern the colonies just before confluence ($\sim 60\%$ full). With the cellular cluster, 80 cells are placed randomly within a much smaller circular area (0.25mm in diameter) representing a small heterogeneous tumor mass shortly after initiation. From this arrangement, the population grows to 8,000 cells before the simulation terminates, which is around the size where vascular recruitment has an effect.  The proximity of cells to viable vasculature, which provides nutrients and oxygen, affects both proliferation and migration. To keep the model as simple as possible, we cut off the simulation around the time that the avascular phase ends. These two spatial configurations allow different degrees of competition due to space constraints (c.f.\ Fig.\ \ref{constraints}a). 

Within each section, we examine different constraints on allowable trait combinations. We first examine each trait separately and then in combination with different constraints on allowable values. The heterogeneity of the cellular population will be investigated for variation in proliferation only, variation in migration only, variation in both traits (all combinations allowed), and the go-or-grow constraint (proliferation rate is inversely mapped to migration rate). Figure \ref{constraints}b displays these constraints in a 2D proliferation-migration phenotype space that we will use to both show the density of combinations and as a coloring scheme representing phenotypes in the spatial distributions.  We conclude with an analysis on how the resulting heterogeneity from these situations contributes to the fitness of the populations as a whole. 

\subsection{Proliferation: different schemes drive different compositions}
\begin{figure}[ht!]
   \centering
           \includegraphics[width=0.95\textwidth]{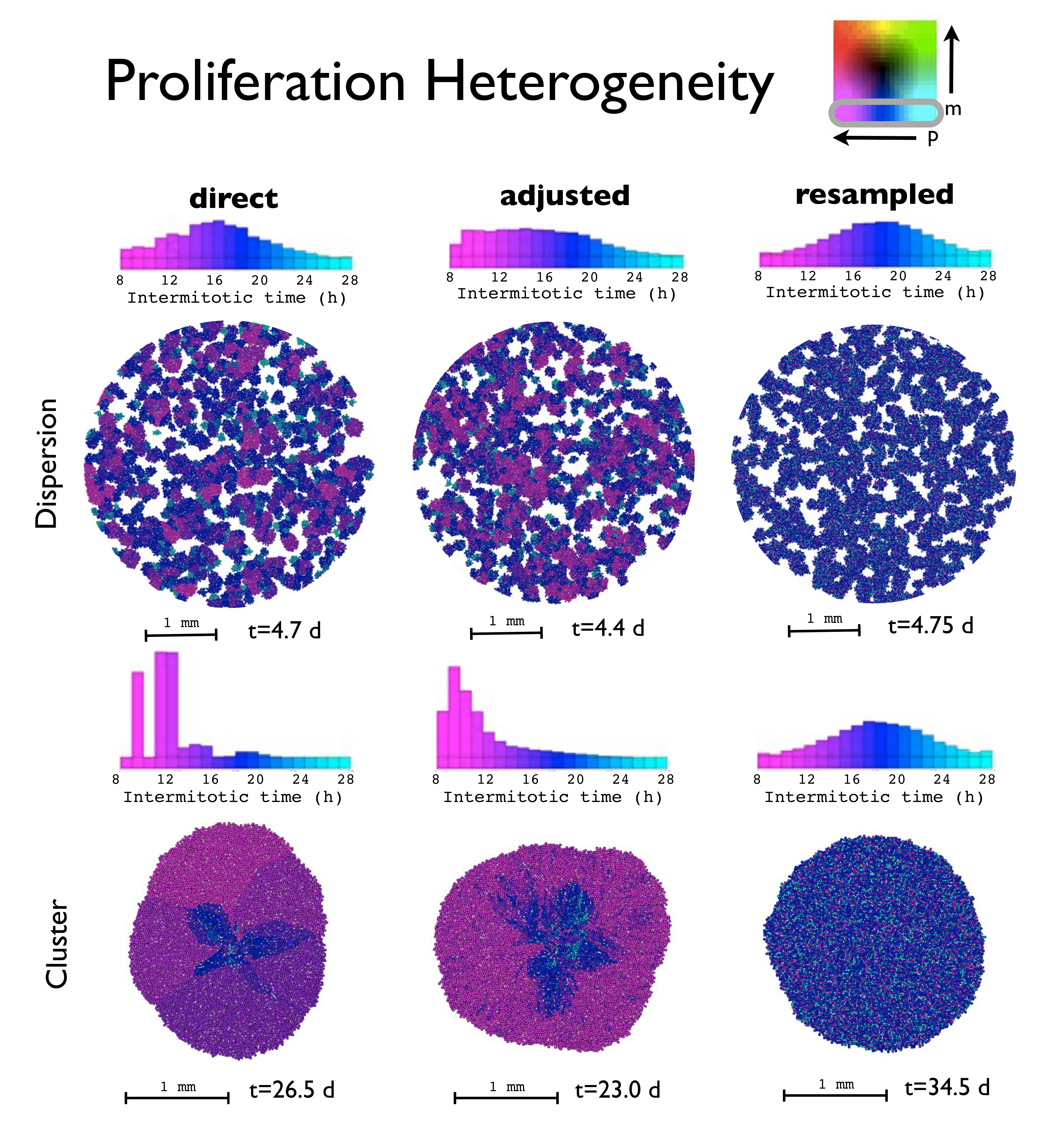}
    \caption{Heterogeneity of IMTs in dispersions upon reaching 25,000 cells (top) and clusters upon reaching 8,000 cells (bottom) with different inheriting modes (columns for direct, adjusted, and resampled inheritance). The spatial distributions are shown with the time taken to reach the final population recorded below. The histograms above the images show the distribution of IMTs at this final population. The gradient from magenta to blue to cyan represents cells with IMTs going from short to long. }%
     \label{pros}
\end{figure}

Cells are given initial IMTs randomly sampled from a normal distribution with a mean of 18 hours and a standard deviation of 4 hours. Trait values are confined within a range of 8 to 28 hours. This is similar to distributions found in real cancer cell lines \cite{Quaranta09}. In Fig.\ \ref{pros} we present the results for nonmotile cells with diversity only in the proliferation rates. 

First and foremost, there is a noticeable difference between the final distributions of traits for the dispersion and the cluster. The dispersion, being spread out spatially, doesn't provoke much selection, but with the extra competition in the cluster, the selection for faster proliferators is much stronger. There are differences in the final distributions of phenotypes across different inheritance schemes. We find that there is selection for faster IMTs if the inheritance is direct or adjusted, whereas no change is observed from the original normal distribution, as expected, with the resampling scheme.  We also see that with direct inheritance, the histogram has more disjointed peaks. With this kind of inheritance if a clone gets stuck in the bulk and goes quiescent, it cannot further propagate its trait. This is a product of both chance and ability. 

The shape and symmetry of the cellular expansion also depends on inheritance mode. For the dispersions, we find that the colonies with short IMTs are larger in the direct and adjusted inheritance modes, while the slower proliferating colonies are smaller. Inheritance resampling, on the other hand, results in colonies of roughly the same size due to an equal diversity in each. Just as we refer to colonies in the dispersion, we will use the term ``families" in the cluster, to signify a group of cells that are all progeny of an initial seed. 

The cellular clusters evolve in space differently due to tougher competition. With direct inheritance, once a family establishes a piece of the proliferating front, it will move outward as a group in radial bands. But at the interface of these patches of families, competition eventually drives the victory of only one. We can see this progression within the structure where some of the bands consisting of slower IMTs taper off next to a faster growing family. In turn this makes the shape lobular at these interfaces. With adjusted inheritance, initial heterogeneity in space may cause a skewed shape to persist for a while, but the cells eventually drift toward being faster at proliferating. The shape of the cluster might reflect the initial heterogeneity but will eventually round out while wobbling slightly around the lower bound of IMTs. With resampled inheritance, the cluster retains a rounded geometry. It can be thought of as being equally heterogeneous locally and globally with a relatively even radial expansion.

\subsection{Migration: movement benefits all}
Next we examine the situation when cells have variation in migration speeds but have the same proliferation rates.  We set all of the IMTs to be 18 hours long, and initialize the cells with speeds drawn randomly from a normal distribution centered at 12 $\mu$m/h and with a standard deviation of 5 $\mu$m/h. We bound the available migration speeds within a window from 0 to 25 $\mu$m/h. This distribution is similar to the values found in \cite{Quaranta09}. 

\begin{figure}[ht!]
   \centering
           \includegraphics[width=0.95\textwidth]{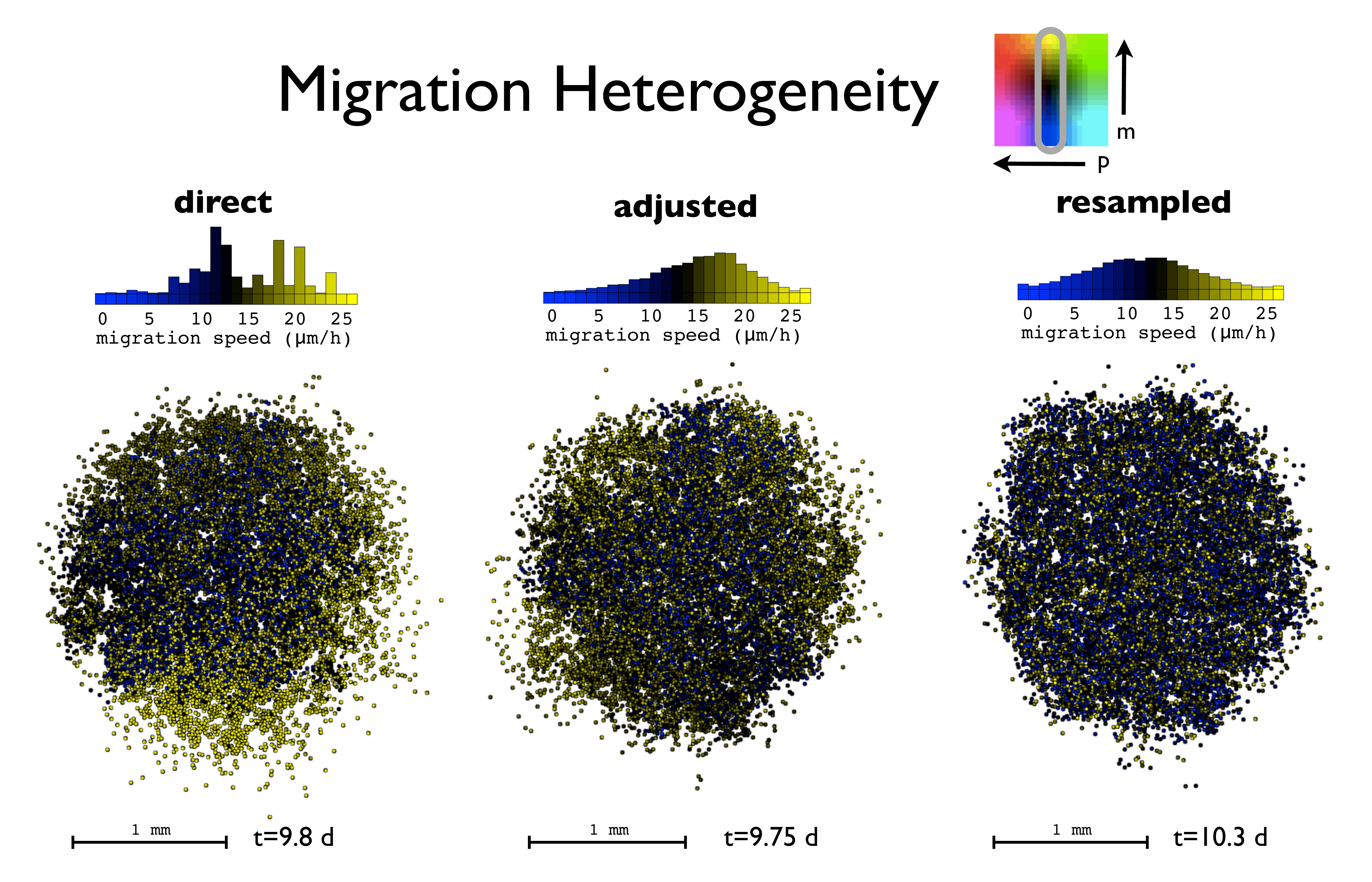}
    \caption{Heterogeneity in migration speeds with different inheriting modes (columns for direct, adjusted, and resampled inheritance). The cells all have the same proliferation rate (18h) but vary in how fast they are moving.  The image shows the spatial distribution upon reaching 8,000 cells, and the time taken to reach this population is recorded below. The histograms above the images show the distribution of migration speeds at this final population. The gradient from blue to black to yellow represents cells with speeds going from slow to fast.}%
     \label{speOnly}
\end{figure}

Using this method with the cellular dispersion, we find that the trait distributions retain their normal shape with no discernible change (not shown). It appears that once the cells can move around, regardless of speed, the selection pressure from spatial competition (that was already small), is almost negligible. There is a more pronounced difference for the cluster configuration with variable migration. These results are shown in Fig.\ \ref{speOnly}. 

We find that there is more competition for space in the cluster configuration, and the faster movers have a selective advantage. However, as they move apart from the mass, it also actually helps the slower movers continue to proliferate by making more space available. So for direct and adjusted inheritance, we see a shift in the distributions over time weighted toward the faster migrators while still maintaining a spread of values as all cells benefit from the outward expansion. We find that similar to the effect of inheritance on proliferation rates, direct inheritance results in a distribution that is skewed toward faster migrators but more rugged and irregular where some clones are snuffed out from proliferating by getting caught up in the quiescent bulk. The adjusted inheritance gives a more smooth distribution of values, and for trait resampling, we find again that the normal distribution is maintained.  

In general, when cells are allowed to move, the whole structure becomes more jumbled, and the proliferating rim widens. There are subtle shape effects on the cluster at the leading edge. For the direct and adjusted inheritance, we see patches where the cells are more spread out and patches where the cells are more packed together. Looking closer, we find that the more diffuse patches correspond, unsurprisingly, to faster movers, and the more compact edges consist mostly of slower moving cells. This effect is more pronounced in the direct inheritance case, where it is more likely that a cell's neighbors are from the same family and have the same trait values. Resampling migration speeds results in an edge that expands rather evenly but that is still moderately diffuse, and a few stray cells may advance farther out due to longer persistence times and/or movement directly away from the mass, creating slight irregularities.

\subsection{Trait combinations: just the sum of the parts?}
\begin{figure}[ht!]
   \centering
    \includegraphics[width=0.87\textwidth]{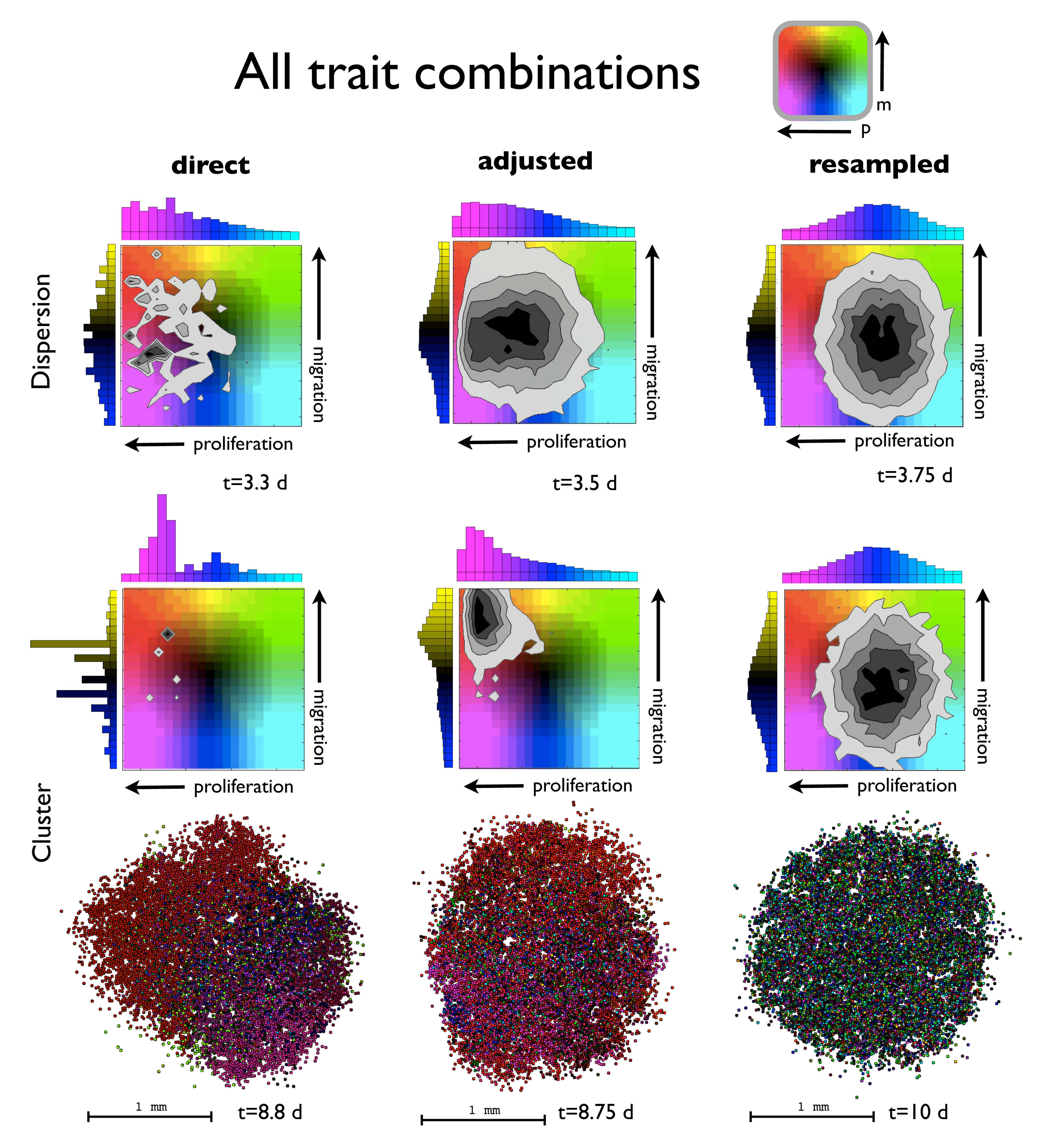} 
    \caption{The frequency of occurrence of cells with combinations of traits in the dispersion upon reaching 25,000 cells (top) and the cluster upon reaching 8,000 cells (bottom). The time taken to reach these populations is listed. The histograms to the top and side of the phenotype density maps show the distribution of proliferation rates and migration speeds, respectively, with increasing values along the direction of the arrow. The phenotype density map represents the frequency of occurrence of each trait combination (darker grayscale values signify more cells with that combination). The spatial layout shows the cluster of cells colored according to where they fall in the phenotype space. 
}%
    \label{bothVary}
\end{figure}

We now take away all constraints on the phenotypes and allow both traits to vary independently of each other. We initialize the cells by randomly sampling from the same normal distributions for IMTs and migration speeds as in the previous sections. We then ran the same sort of simulations as before. The final distributions of traits are shown in Fig.\ \ref{bothVary}. 

Comparing the histograms from Figs.\ \ref{pros} and \ref{speOnly} where each trait was varied on its own to here where both traits vary independently of each other, we notice that there is very little difference between the distributions. The dispersions still do not invoke as much selection as the clusters. The direct and adjusted inheritance schemes select for both faster proliferators and faster movers, but the resampled scheme maintains its distribution. The direct inheritance has disjointed peaks as only several clones dominate, and the adjusted inheritance is smooth over more values. Once again, faster proliferators are more strongly selected for than faster movers. As far as the distributions in traits is concerned, the combination of traits appears no different than the sum of the two individual traits.   

Because of the greater competition and better image resolution, we show only the spatial results for the cluster configuration, and though the mass appears somewhat jumbled, we can still pick out some spatial trends. For both the direct and adjusted inheritance, there is some clumping together of similar traits (red and magenta patches), but the clumping is more apparent with direct inheritance. We also notice that some slower proliferators are speckled throughout (black and green cells).  Like in the previous section, it is mainly the fast proliferators that get to the edge, but so do some slow proliferating, fast migrating cells. The shape of the mass with direct inheritance is elongated where the faster proliferators have dominated the region.  Where the faster migrators have reached the leading edge, the proliferating rim is more diffuse.  The resampled inheritance leads to a mix of cells with normally distributed trait values for both traits.  The cells reside in a relatively even spatial distribution throughout the mass, and the shape is quite round compared to the others. 
 
The traits don't appear to be correlated. Even though there is a complicated spatial distribution for the traits, it is clear that the combinations of traits that get selected when both traits vary are quite similar to the combination of independent traits that get selected when each trait is varied separately. However, the story changes when we limit the phenotype space by imposing a trade-off between proliferation and migration.  

 \subsection{Go-or-grow: better do both}
So far we have considered the traits of proliferation and migration as separate and also together but uncorrelated, but this may not be so. Rapidly advancing through the cell cycle and synthesizing new material for division and actively moving around may cost the cell a large amount of energy. The cell may only have so much energy to expend, so we consider the idea of a trade-off between proliferation and migration, often referred to as go-or-grow. We assume that if a cell is a faster migrator, it must be a slower proliferator. The migration speed is mapped linearly to the proliferation rate along the diagonal axis of the phenotype space in this manner. 

\begin{figure}[ht!]
   \centering 
    \includegraphics[width=.87\textwidth]{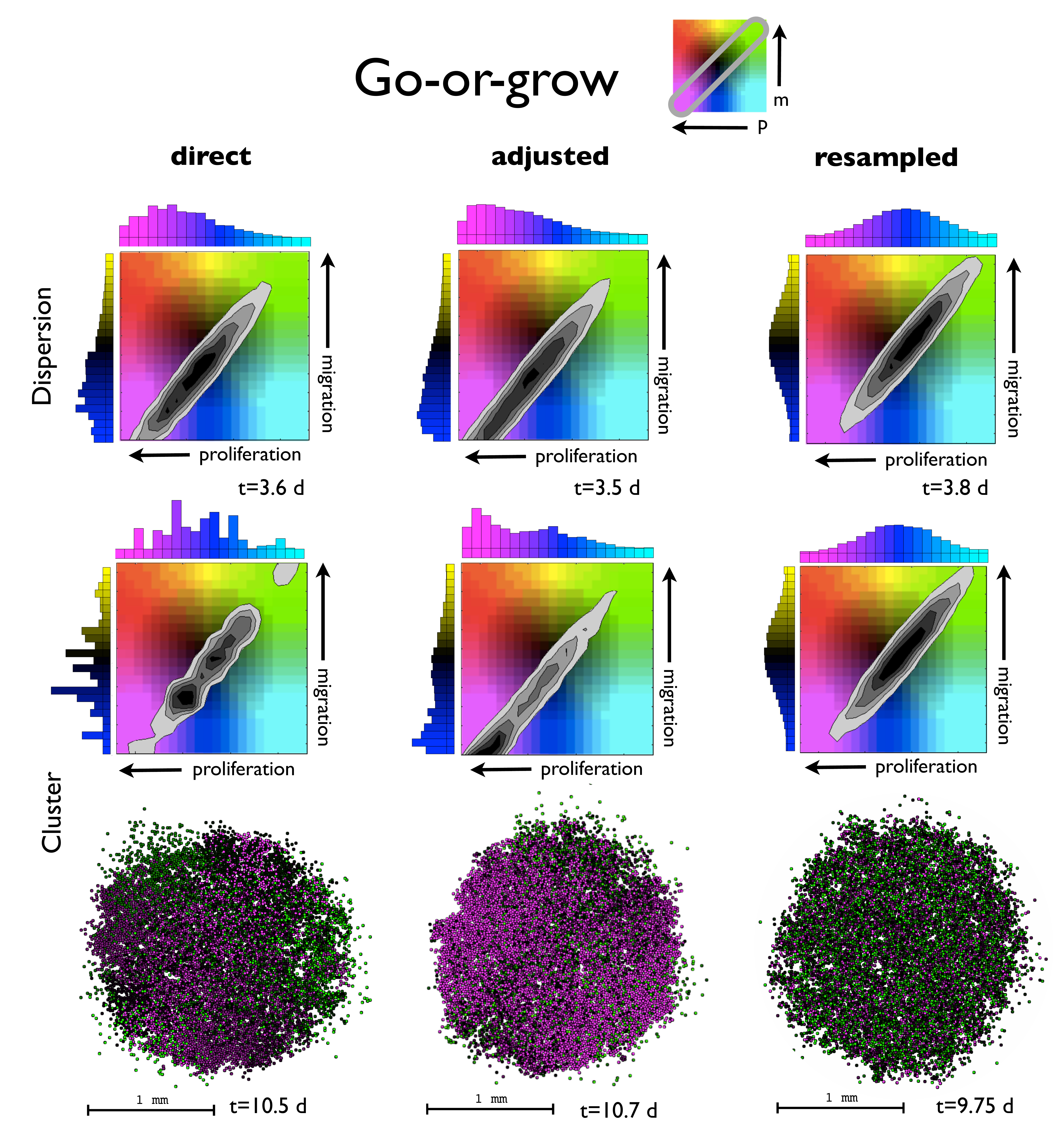} 
    \caption{The frequency of occurrence of cells when confined to the diagonal of the phenotype space (go-or-grow) in the dispersion upon reaching 25,000 cells (top) and the cluster upon reaching 8,000 cells (bottom). The time taken to reach these populations is listed. The histograms to the top and side of the phenotype density map show the distribution of proliferation rates and migration speeds, respectively, with increasing values along the direction of the arrow. The phenotype density map represents the frequency of occurrence of each trait combination (darker grayscale values signify more cells with that combination). The spatial layout shows the cluster of cells colored according to where they fall in the phenotype space.}%
 \label{GGAll}
\end{figure}

The resulting distributions of traits are shown in Fig.\ \ref{GGAll}. We see once again that a shift in trait distributions occurs with the direct and adjusted inheritance, whereas trait resampling results in practically no change from the original distribution. For the dispersion, there is a slight shift toward faster proliferating cells, which means that there is a slight shift toward slower migrating cells. This is not surprising with the cells spread out like this, because faster proliferators are more strongly selected for than the faster migrators when the cells already have enough space. But for the cluster, something interesting happens. It is most apparent in the adjusted inheritance scheme. The same overall shift occurs toward faster proliferators that are slow at moving, but another population appears to also be present. In addition to the peak at fast proliferation values, there is another peak in the middle of the distribution where the cells are average at both proliferating and migrating. For direct inheritance, there may be multiple peaks, but with the inherent bumpiness of the distribution it is hard to decipher. 

The spatial distributions of the clusters are presented in Fig.\ \ref{GGAll} for the three types of inheritance. We find that in general, the assemblage is more heterogeneously mixed, than with the other constraints. With direct and adjusted inheritance, there are some regions with similar neighbors, but there are also regions that are more mixed. The resampled inheritance again maintains its usual diversity. 

\begin{figure}[ht!]
   \centering 
    \includegraphics[width=0.95\textwidth]{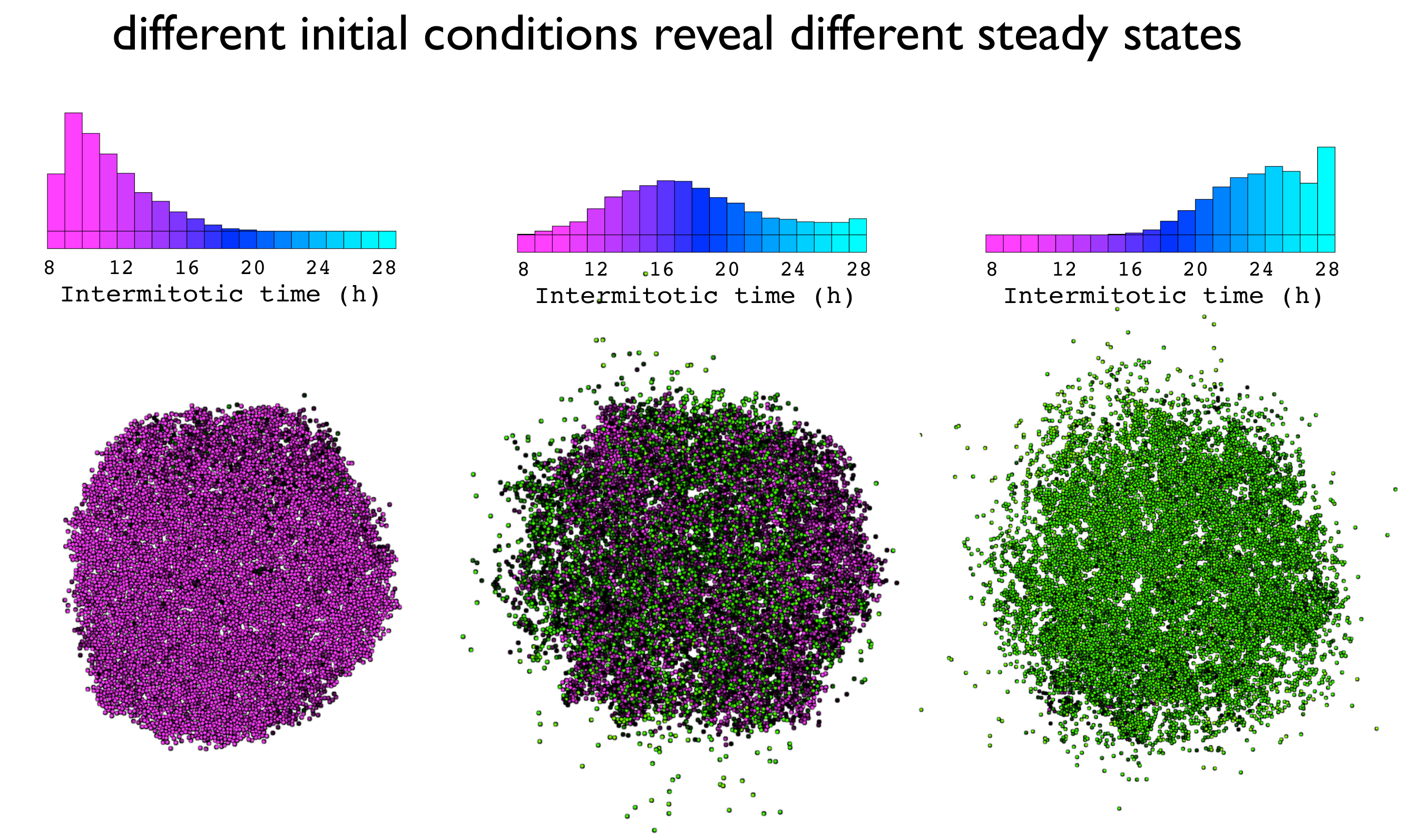} 
    \caption{With the adjusted inheritance, the go-or-grow constraint yields different populations of clusters grown to 8,000 cells depending on the initial distribution. All initial distributions are monoclonal with IMTs at 10 hours (left), 18 hours (middle), and 26 hours (right).}%
 \label{multimod}
\end{figure}

Further investigation of the adjusted inheritance for the cluster configuration reveals that the weak bimodality is actually quite unstable. The peak may either stay in the middle, move toward faster proliferators, or become bimodal with both peaks present to some degree as previously described and shown in Fig.\ \ref{GGAll}. The initial distribution has a significant effect on where this final distribution will stabilize. Figure \ref{multimod} shows the final configurations for simulations starting with all cells either fast at proliferating, fast at migrating, or average at both. In these cases, the distribution tends to remain centered around where it began. However, when the trait values begin in the middle of the range, several outcomes are possible: either the peak remains in the middle, another peak forms at fast or slow IMTs, or several peaks arise.

The appearance of a multimodal or bimodal distribution when the inheritance is adjusted most likely means that many trait values are equally fit. Since the resampled scheme, which is very heterogeneous, grows quicker, the indication is that local heterogeneity is necessary for maximal growth. Local heterogeneity is harder to achieve when the inheritance has some familial memory (direct and adjusted inheritance gives rise to daughter cells exactly like or similar to their parent, respectively), but when neighbors can easily be different from each other, as in trait resampling, the local heterogeneity is most prevalent. Therefore, though a single cell has restrictions on which combinations of traits it can have (go-or-grow), local neighbor interactions of mixed phenotypes can lead to unexpected cooperation (where it may do both). 

\subsection{The heterogeneity index: quantifying the phenotypic mix}
So far we have shown how different constraints on the phenotype space in heterogeneous tumor growth can affect cell population heterogeneity both temporally and spatially. To understand the trait variation and allotment within a range of values, we have plotted histograms of single traits. To understand the variation and abundance of trait combinations, we have shown phenotype density maps. Now we want to compare the previous results by describing the heterogeneity numerically. We use the heterogeneity index outlined in Appendix 1 as our metric for comparison.
 
\begin{figure}[ht!]
   \centering
    \includegraphics[width=0.9\textwidth]{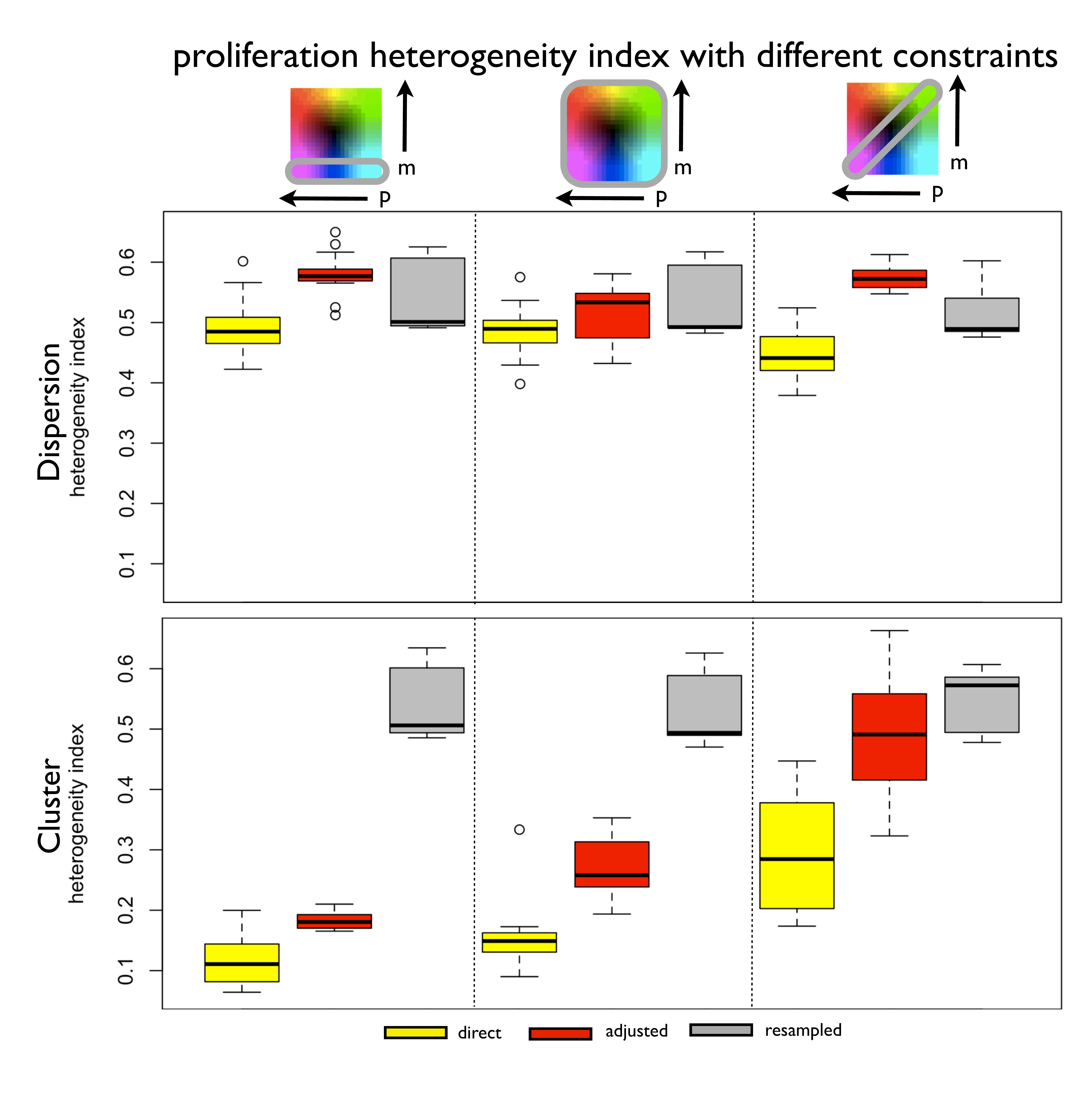} 
    \caption{Heterogeneity in proliferation rate as calculated via  Eq.\ \ref{hetIndexEq} for each constraint on phenotype space (columns), for each inheritance type (colors), and each spatial configuration (top is the dispersion and bottom is the cluster). The columns are for: proliferation only, both traits with no constraints, and go-or-grow. From fifteen different runs, the range, upper quartile, and lower quartile are shown.}%
        \label{HETP}
\end{figure}

\begin{figure}[ht!]
   \centering
    \includegraphics[width=0.9\textwidth]{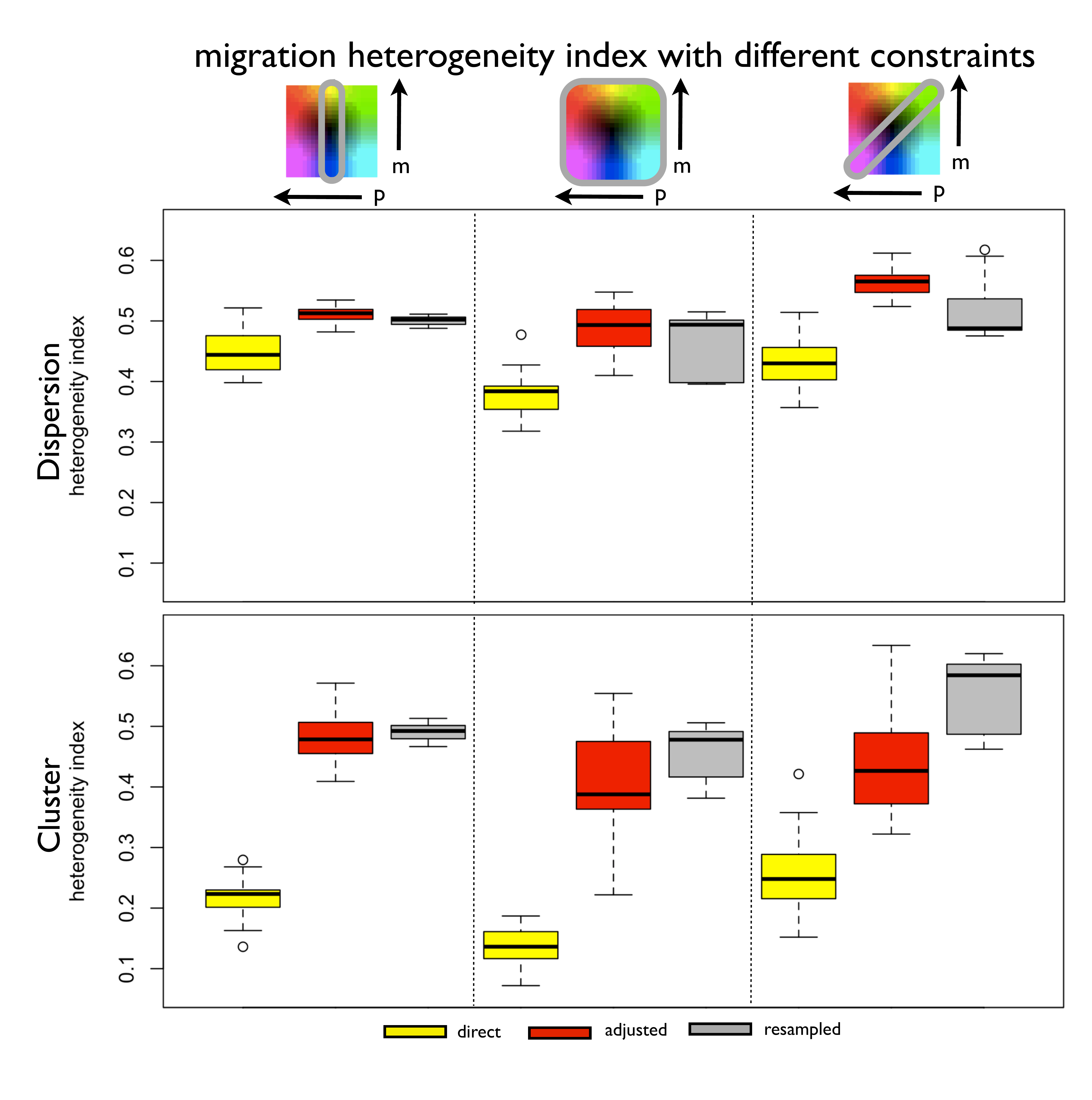} 
    \caption{Heterogeneity in migration rate as calculated via  Eq.\ \ref{hetIndexEq} for each constraint on phenotype space (columns), for each inheritance type (colors), and each spatial configuration (top is the dispersion and bottom is the cluster). The columns are for: migration only, both traits with no constraints, and go-or-grow. From fifteen different runs, the range, upper quartile, and lower quartile are shown.}%
        \label{HETM}
\end{figure}

In the previous sections, we showed typical results for each scenario, however, with the amount of stochasticity involved in these simulations, variation is expected. Here we compile from 15 different simulations of each scenario the heterogeneity indices for proliferation rates (c.f.\ Fig.\ \ref{HETP}) and for migration speeds (c.f.\ Fig.\ \ref{HETM}). The heterogeneity index $H$ was calculated from the final trait distributions.  

The mean heterogeneity index is generally higher in the dispersions where there is less selection, than in the clusters, but there are also differences between the different inheritance modes. The direct inheritance scheme consistently yields the smallest mean $H$ across all scenarios. This is to be expected, as no new traits can be gained but only lost. The resampled scheme appears to remain relatively steady over the different constraints, while the mean values of $H$ for the adjusted inheritance scheme jump around to different values.  

For the cluster, the mean $H$ always increases from direct to adjusted to resampled inheritance, and the extra competition in the cluster configuration makes the homogeneity from direct inheritance more pronounced. The first two columns reflect that the mean values for both the direct and adjusted modes move toward faster proliferation, so this results in a distribution that is more homogeneous. However, with go-or-grow, many values are equally fit, so there is not one trait value to move toward. This results in more heterogeneity, and also more variation in the amount of heterogeneity as there are many outcomes possible.

When comparing the heterogeneity indices for migration, similar trends are present, but there are differences. Again, direct inheritance has the lowest mean value of $H$, but resampling does not always have the the highest mean $H$. With migration, the adjusted scheme will often tend toward a very spread out distribution, which translates to a large $H$. But the heterogeneity index is largest in the go-or-grow scenario with the resampling scheme. Now, with a good handle on heterogeneity in this system, we define how it leads to different population fitnesses.

\subsection{Fitness: different fates for different constraints}
The heterogeneity of a population and its distribution of phenotypes gives some measure of its capacity to survive in different environments, but there are other ways to characterize the population's fitness. The most obvious metric is the overall growth rate of the population, which we will evaluate. We will also look at the proportion of proliferators in the population, which gives a measure of spatial diffusiveness, a degree of competition from neighbors, and an overall capacity for continued proliferation. 

\begin{figure}[ht!]
   \centering
           \includegraphics[width=0.95\textwidth]{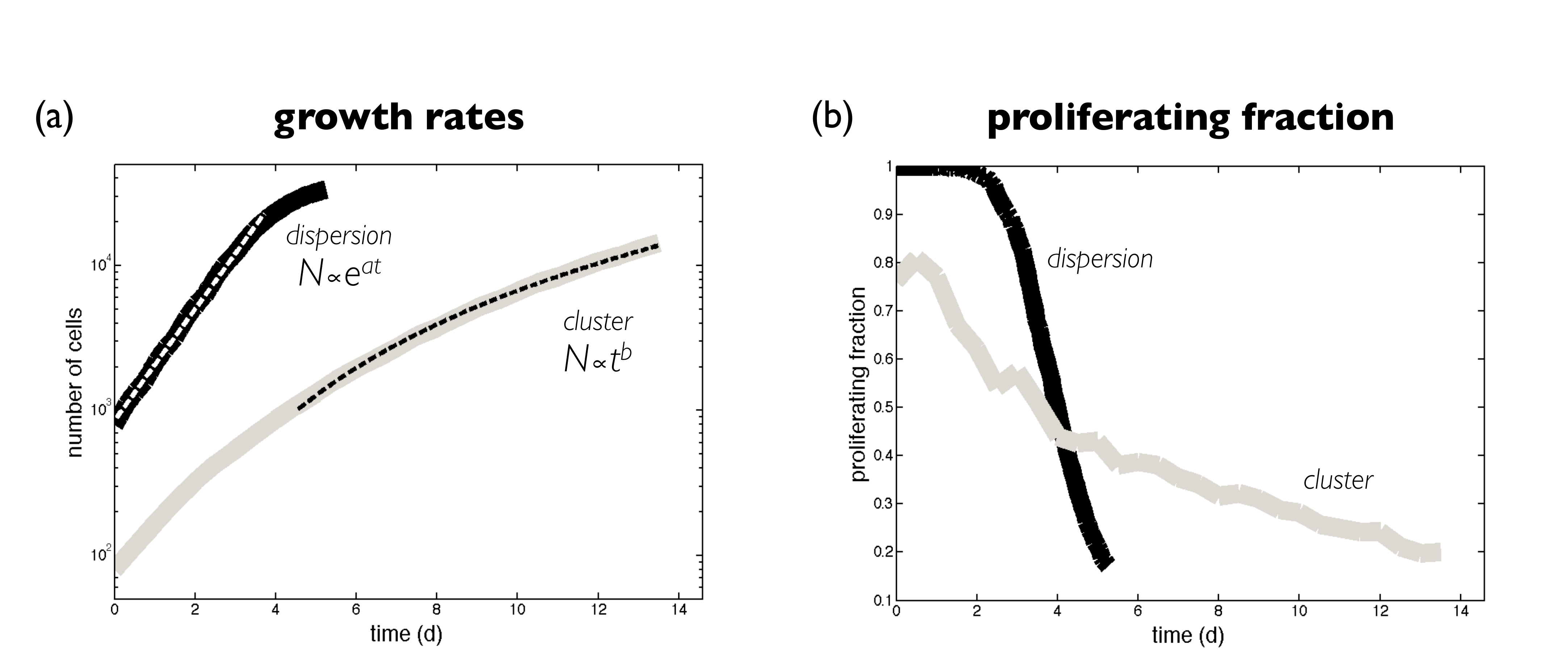}
    \caption{a) Typical growth rates of the dispersion and cluster configurations. The dispersion is best fit with an exponential, whereas the cluster is best fit to a power law (both with $R^2>0.999$). b) The percentage of cells of the whole population that are in the proliferating state. The two spatial configurations clearly lead to different degrees of competition.}%
\label{GRPQ}
\end{figure}

The growth rates are quite different for the dispersion and the cluster (c.f.\ Fig.\ \ref{GRPQ}a). The dispersion grows exponentially ($N \propto e^ {at}$) until reaching confluence, and then it levels off as the space eventually fills to capacity. For the cluster configuration, the cells do not immediately compete with each other as there is some space between them from the random nature of the initial placement. So after a quick growth spurt when the population is small, the mass gets larger and the fraction of cells at the leading edge that can proliferate reduces. A power law ($N \propto t^ {b}$) fits the growth of the cluster best. 

The proliferating fraction for the dispersion and the cluster have different temporal dynamics (c.f.\ Fig.\ \ref{GRPQ}b). For the dispersion, all cells start their own colonies with 100\% in the proliferating state. Each colony grows initially like its own separate cluster building up individual quiescent cores, but quickly the space fills up and neighboring colonies become the competition. The initial proliferating fraction for the cluster is smaller than the dispersion. This fraction shrinks very gradually as the quiescent tumor bulk becomes a larger and larger piece of the whole population compared to the proliferating rim.

Understanding how the fitness is affected by different inheritance schemes and different constraints on heterogeneity is not straightforward. We ease into this intricacy by first looking at the fitness of trait combinations in homogeneous populations. For simplicity and to be able to compare the dispersion with the cluster, we correlate fitness with the growth rate of the population, which we take as $(1/N_0)(dN/dt)$, where $N_0$ is the initial population, $dN$ is the change in total number of cells and $dt$ is the total time.

\begin{figure}[ht!]
   \centering
    \includegraphics[width=0.9\textwidth]{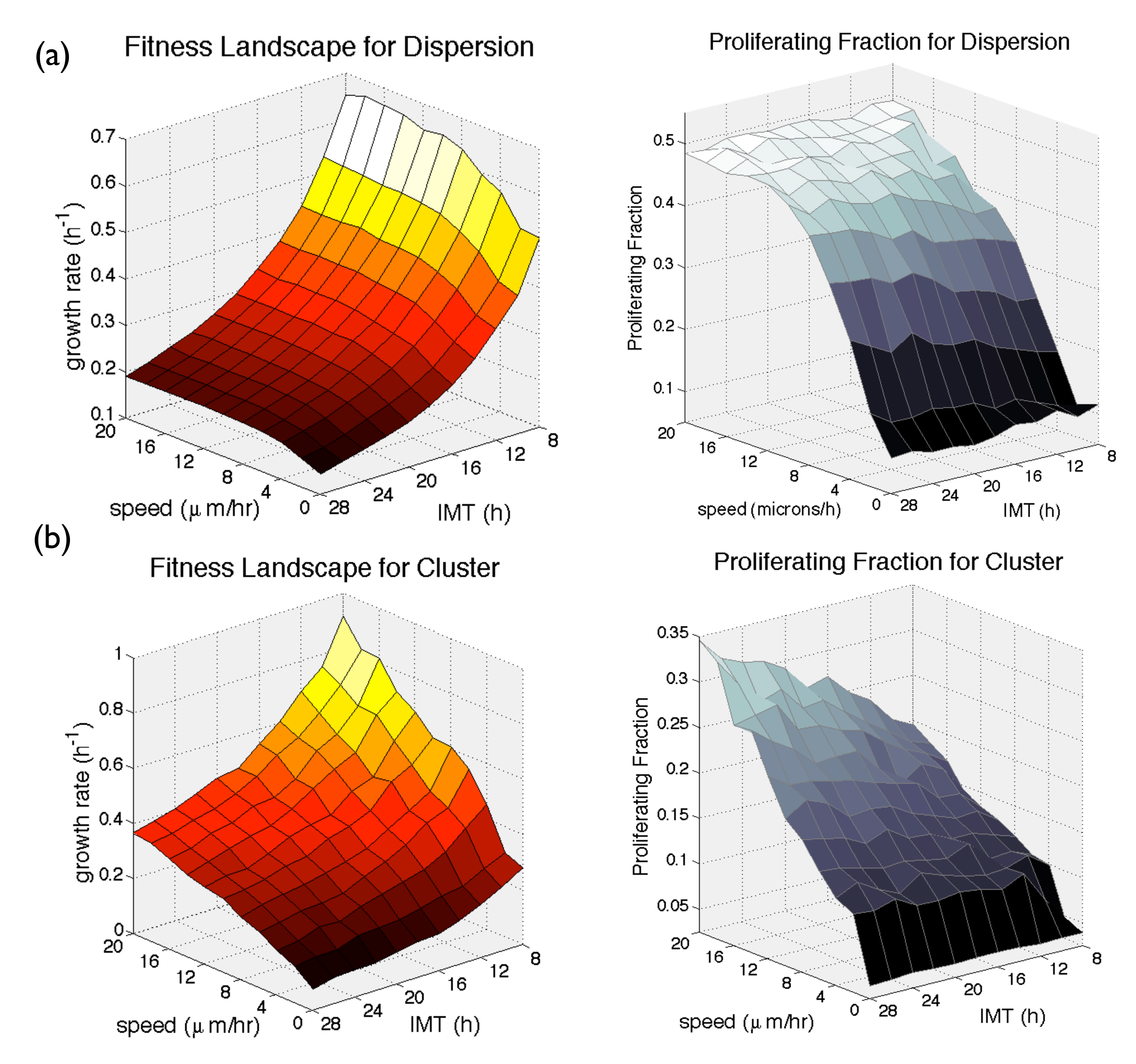} 
    \caption{Growth rates (colored) and proliferating fractions (grayscale) of monoclonal populations with various combinations of proliferation and migration rates. To quantify the growth rate and proliferating fractions, a) the dispersion was grown from 800 to 25,000 cells, and b) the cluster was grown from 80 to 8,000 cells. The growth rates are found by dividing the total change in the number of cells by the total change in time. This is then normalized by the initial population. The proliferating fraction is taken at the final time step.}%
\label{mono}
\end{figure}

We grow monoclonal populations (all cells have identical trait values which are passed on directly to their progeny), and record the overall growth rate. We take the proliferating fraction at the end of each simulation. The results are shown in Fig.\ \ref{mono} as fitness landscapes over the 2D space of phenotype combinations.  

For the cellular dispersion in Fig.\ \ref{mono}a, the growth rate surface plot indicates that being a faster proliferator makes the biggest impact on fitness. Increasing the speed is shown to increase the proliferating fraction, but it does little to increase the overall growth. However, there is a noticeable decline in the growth rates at very slow migration speeds. In this configuration, the cells are already proliferating exponentially, so allowing them to spread out hardly makes a difference. 

For the cellular cluster in Fig.\ \ref{mono}b, the surface plot shows that the most fit combination of traits corresponds to the fast proliferating, fast migrating phenotype, and the least fit combination corresponds to the slow proliferating, slow migrating phenotype. This agrees with what we find to be the most fit when there is heterogeneity in the traits - the population trends toward this most fit combination. However, we also see that there are many combinations in between these extremes that share the same values for fitness. From this figure we find that upon the line allowed by the go-or-grow condition, most values are equally fit. The combination with the largest proliferating fraction, however, does not correspond to the fastest growth rate. The largest proliferating fraction occurs when the cells move out fast, but proliferate slowly. This combination creates a very diffuse tumor that invades farther before filling in the space left behind.

Analysis on monoclonal populations only tells us so much. It is not always the fastest moving and most proliferative populations that come out on top when many interacting parts make up the whole. Just as certain combinations of traits may work well together, combinations of neighbors may interact differently. Furthermore, the fitness of the population as a whole cannot be determined by summing the fitnesses of the clones that make up the population, as we see next.

\begin{figure}[ht!]
   \centering
    \includegraphics[width=0.95\textwidth]{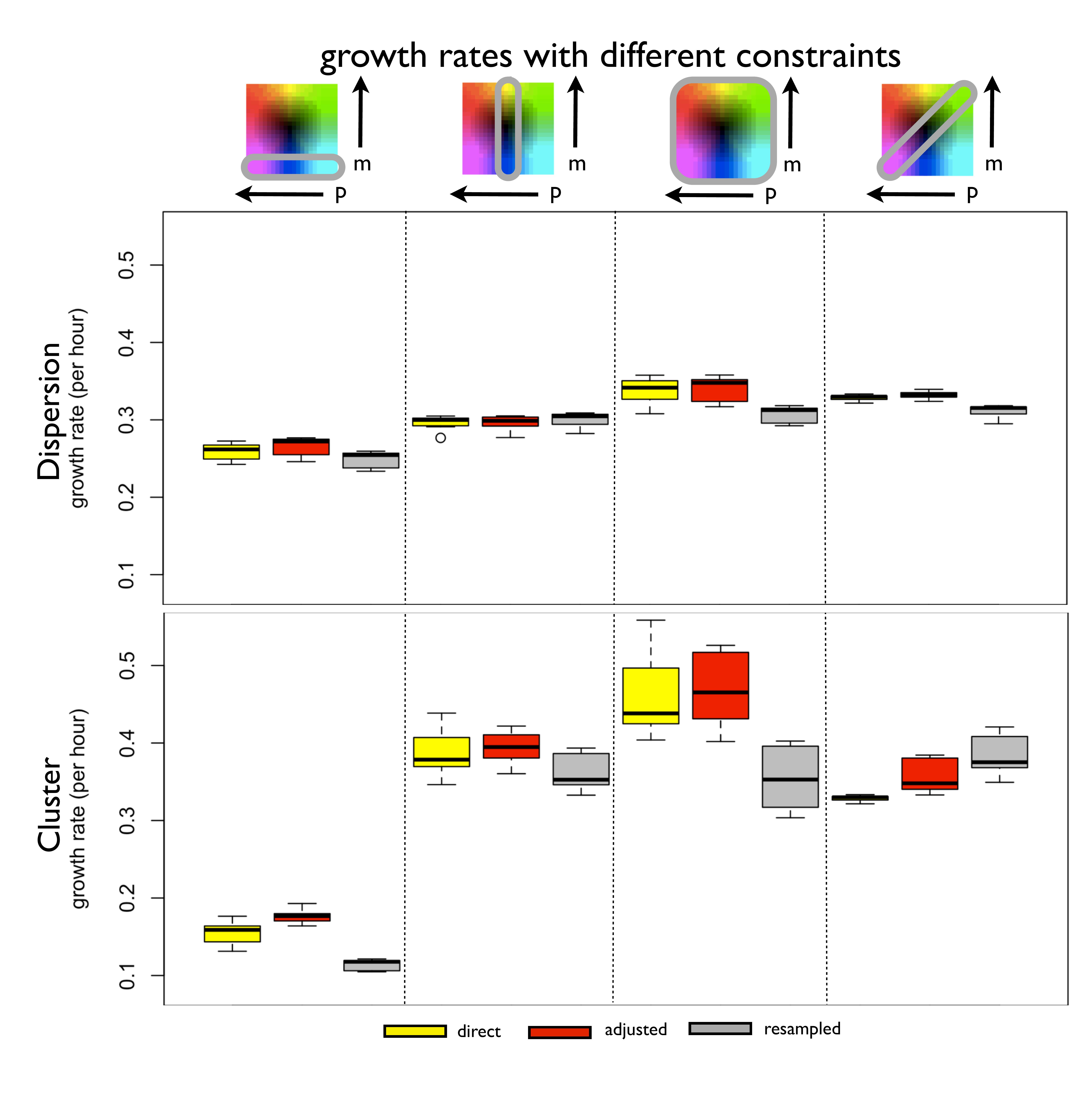}
    \caption{
    Growth rates for the dispersion (top) and the cluster (bottom) for each constraint on phenotype space (columns) and for each inheritance type (colors). From 15 different runs, the range, upper quartile, and lower quartile are shown.}
\label{GRAll}
\end{figure}

We now address the complexity of heterogeneity from the previous sections. We compare the fitness metrics of each of the different inheritance schemes with different constraints and different spatial configurations by compiling results from 15 simulations for each scenario. First we look at the fitness of the populations, measured by the growth rate (c.f.\ Fig.\ \ref{GRAll}).   

For the dispersion configuration, there is less difference between the three inheritance schemes, however, there is a tendency for the resampled inheritance to yield slower growing populations compared to the adjusted and direct inheritance schemes. But when migration varies as a single trait, increasing migration speeds (as seen with direct and adjusted inheritance) does little to help when there is already such meager competition, so there is hardly any advantage to select, and the three inheritance schemes produce similar results. 

The growth rates for the cluster configuration are more distinct between inheritance schemes. In the first three columns, the adjusted and direct inheritance schemes yield faster growth rates than the resampled, which lags behind. But in the last column, for go-or-grow, the resampled scheme's mean growth rate is faster. In the first three columns, we know that the distribution tends toward homogenous populations with faster proliferators and faster migrators. But when these values are no longer allowed and many trait values are equally fit (in go-or-grow), this heterogeneity that previously reduced the fitness of the population, is now beneficial. Though the adjusted and direct inheritance schemes are seen to also lead to increases in the heterogeneity indecies for both traits for go-or-grow, these schemes will never achieve the same degree of local heterogeneity that the resampled scheme is capable of providing, which is presumably a growth advantage.  

\begin{figure}[ht!]
   \centering
    \includegraphics[width=0.95\textwidth]{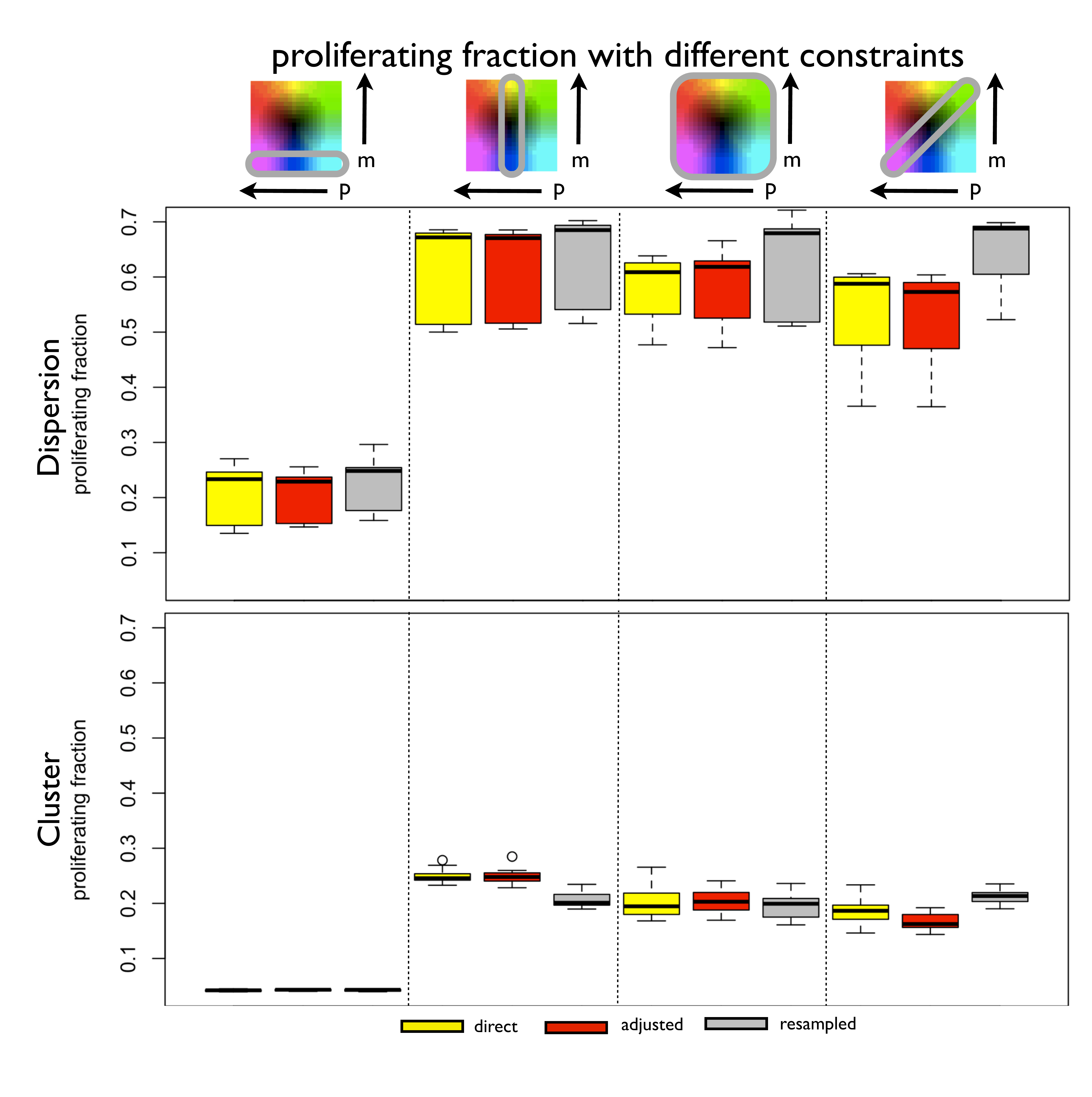} \caption{The percentage of the total population of cells that are proliferating for the dispersion (top) and the cluster (bottom). Fifteen different runs make the variance in the plot. }
\label{PQAll}
\end{figure}

Fitness is not just a measure of the growth rate of a population. A population of cells may also be more fit if there is a larger proportion of cells in the proliferating state. So we examine for all situations, the percentage of proliferators of the total population at the end of each simulation (c.f.\ Fig.\ \ref{PQAll}). 

We record the proliferating fraction for the dispersion when it is around $60$\% confluent, whereas the cluster's proliferating fraction is measured when it reaches a size of around 2 mm in diameter. Therefore, the proliferating fraction is always larger in the dispersion than in the cluster due to the fact that we are comparing the edges of many proliferating colonies with the edge of one big proliferating mass, respectively. Also, when the cells don't migrate (the first column), the mean fraction is always smaller, because there is no diffuse boundary of cells at the edges that increase this proliferation ratio value. Once the cells have some movement, the edges become more diffuse, and this value increases.

For the dispersion, the mean fraction of proliferators for the resampling inheritance mode is larger than the others, and this is more pronounced in the go-or-grow scenario. For the cluster, the resampling scheme actually appears to have mean proliferating fractions that are less than or equal to the other schemes, yet the value once again is largest for the go-or-grow scenario. The drift toward faster migrators with direct or adjusted inheritance could lead to a diffuse boundary, but the local heterogeneity found with the resampled inheritance could also lead to a diffuse edge  by avoiding large pockets of closely packed fast proliferators.

We have examined three metrics to numerically evaluate the heterogeneous population growth of this system: the heterogeneity index, the growth rate, and the proliferating fraction. Not one of these metrics gives the a complete understanding of the dynamics, but each has provided some insight into piecing together how the many interactions between the parts leads to the dynamics of the whole.
 
\section{Discussion}
This analysis has brought to light that with heterogeneity in traits, different modes of inheritance of phenotypes can result in different outcomes as far as population composition, heterogeneity, and fitness. The only environmental constraint considered here is space, and lack of space can drive selection toward populations that are more or less heterogeneous depending on the allowed trait combinations.   

An obvious question to ask is, is one of these inheritance schemes more realistic? It is probable that some traits don't change often upon division like with direct inheritance, and that some adaptability is also allowed like the adjusted inheritance. The unchanging diversity of the resampling scheme seems quite unlikely, yet we saw in the go-or-grow scenario that this local heterogeneity provides an advantage to the growth of the population, and it also might be beneficial when first entering a new microenvironment (e.g.\ a new metastatic niche) to be maximally heterogeneous to better evaluate which traits work best. With all of the ways that function actually comes about in a cell, it's also not necessary that every trait will be passed on in the same way. 

\begin{figure}[ht!]
   \centering
           \includegraphics[width=0.9\textwidth]{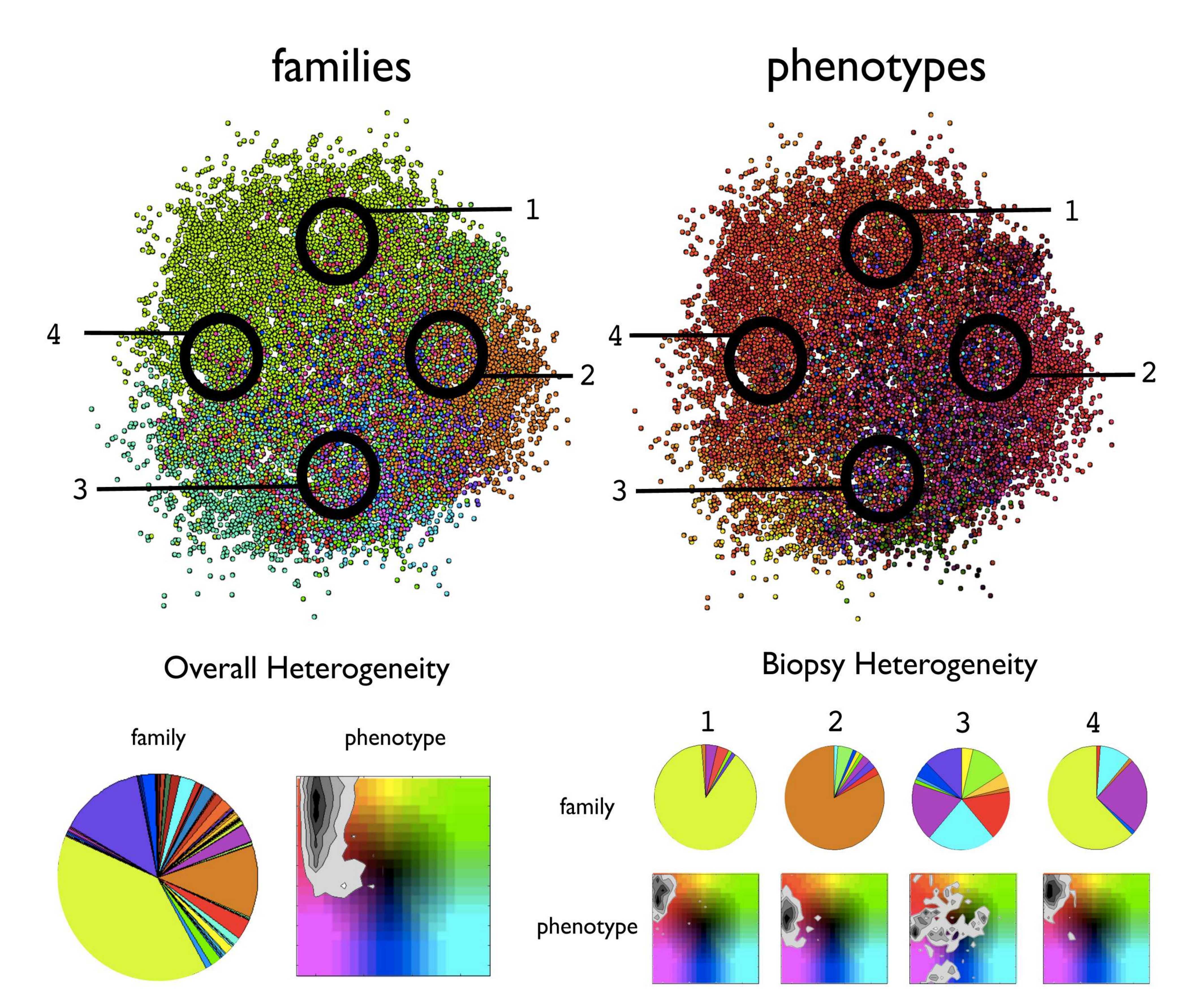}
    \caption{Heterogeneity from familial origin (left) and phenotype (right) does not always correspond. Site-specific differences are also found. We see that even if a tumor may have several dominant clones, the phenotypes may converge.}%
   \label{biopsies}
\end{figure}

It is clear that with an inheritance scheme that allows plasticity of a phenotype, the trait may converge to be more homogeneous even though they came from different clones of origin. This idea of convergent evolution is not new, but has recently been demonstrated by genetically characterizing biopsies from different sites within a tumor \cite{Gerlinger12,Sottoriva13}. To explicitly confirm this phenomenon in our model, we use the scenario where the trait combinations are unlimited and the inheritance is adjusted (c.f.\ Fig.\ \ref{biopsies}).

We color the images in Fig.\ \ref{biopsies} according to ``families", or progeny from the same clone of origin (left), and position in the 2D phenotype space as before (right). We see that each biopsy site results in different distributions of families and different distributions of phenotypes. Further, the originating family and phenotype need not be related. Looking at the overall heterogeneity, it is clear that there are multiple families producing the same phenotypic behavior (the phenotype density map shows convergence towards faster proliferation and faster migration). If we then look at each biopsy in isolation then this pattern is similar for biopsies 1, 2, and 4. In biopsy 3, we see that the families and phenotypes are similarly diverse. This na\"{i}ve example shows how looking at the genotype (or here, the original clonal family) might be misleading, and as these tumors grow the divergence between genotypes will grow while the convergence to phenotypes will also increase. It is worth noting that there is still utility in looking at the spatial distribution of genotypes as those that dominate will be correlated with the fitter phenotypes. However, our example here has only 80 initial ``genotypes" that do not mutate and is certainly a gross underestimation of reality. We therefore believe that it is critical to characterize the phenotypic heterogeneity (at the single cell level) even more than the genotypic heterogeneity as ultimately this will facilitate our understanding of how the tumor grows and might be better treated.

Advances in single cell analysis have allowed us to examine more fully the heterogeneity of genetic and phenotypic states of individual cells \cite{Quaranta09,Rowat09b}. But maybe more important than understanding the variation in single traits is the heterogeneity of certain trait combinations and how they are distributed spatially. For just the two traits we have examined here, there is the potential for one trait to influence the other. The faster a cell divides the faster the whole population can grow, but if it is not moving apart it only keeps a very small proliferating rim so that the growth is limited. There is a balance needed for optimal growth between keeping the cells dividing to grow the population and moving out to keep the cells in the proliferating state. Beyond just looking at growth rates for a measure of fitness, the proportion of the population that is proliferating is significant. 
 
We incorporate competition for space as the only environmental factor (in both the dispersion and the cluster configuration), which provides a very simplistic view of proliferation and invasion of a population. Nevertheless, an important feature of these two very different spatial configurations is that the growth, evolution, heterogeneity, and fitness of the populations are dramatically different. We should be careful in interpreting in vitro models where the cells do not experience this competition or make more observations when cells are closer to confluence. We know that in vitro models are, nonetheless, models and should be considered as such. Biological models like tumor spheroids, scratch assays, 3D cultures, and the Nest assay \cite{Kam09,Kam11} are better at mimicking spatial competition over standard cell cultures. These models represent the migration much more realistically, but with so many combinations of traits, there is no way to study them all without the help of mathematical and computational models.

Some scenarios played out here may be more realistic than others though all are quite abstract from the complexity of a real tumor. Though abstract, this method of heterogeneity analysis of trait inheritance could be played out with any combination of traits. We could also consider the evolution of populations with more than two traits as angiogenesis, metabolism, build-up of acidity and other processes change the microenvironment. The complexity with just two traits is already quite extensive, but a multidimensional phenotype space is also possible. With each additional trait considered, the combinations of traits grows quickly, but realistic constraints on these combinations may significantly reduce the dimension as in the go-or-grow scenario. 

The spatial and temporal evolution of phenotypic heterogeneity (as defined by trait combinations) in a tumor has significant implications for the treatment of that tumor \cite{Swanton12}. With the growing trend of genetic characterization of tumors from a single biopsy, we may only be capturing a skewed subset of the whole heterogeneity. Beyond the potential for misconstruing the type of treatment based on where the biopsy is obtained, when a certain clone or phenotype is targeted, we may be freeing up space and resources for other cell types to take over. If we look at tumor growth and regrowth during and after treatment as an evolutionary process, we may get a better handle on how to best prevent recurrence. The mode of phenotypic inheritance directly affects tumor growth, and the interplay between the various traits is a source of complexity that deserves much more attention from the biological community as this ultimately may be the place where the best therapeutic strategies for targeting a given tumor are found. 

\section{Acknowledgements}
The authors gratefully acknowledge funding from the NCI Integrative Cancer Biology Program (ICBP) grant U54 CA113007.

\bibliography{bibfile.bib}

\begin{thebibliography}{10}

\bibitem{Anderson05}
A.~R.~A. Anderson.
\newblock A hybrid mathematical model of solid tumor invasion: the importance
  of cell adhesion.
\newblock {\em Mathematical Medicine and Biology}, 22:163--186, 2005.

\bibitem{Anderson09}
A.~R.~A. Anderson, M.~Hassanein, K.~M. Branch, J.~Lu, N.~A. Lobdell, J.~Maier,
  D.~Basanta, B.~Weidow, A.~Narasanna, C.~L. Arteaga, A.~B. Reynolds,
  V.~Quaranta, L.~Estrada, and A.~M. Weaver.
\newblock Microenvironmental independence associated with tumor progression.
\newblock {\em Cancer Research}, 69:8797--8806, 2009.

\bibitem{Anderson06}
A.~R.~A. Anderson, A.~M. Weaver, P.~T. Cummings, and V.~Quaranta.
\newblock Tumor morphology and phenotypic evolution driven by selective
  pressure from the microenvironment.
\newblock {\em Cell}, 127:905--915, 2006.

\bibitem{Hatzikirou12}
K.~B\"{o}ttget, H.~Hatzikirou, A.~Chauviere, and A.~Deutsch.
\newblock Investigation of the migration/proliferation dichotomy and its impact
  on avascular glioma invasion.
\newblock {\em Math. Model. Nat. Phenom.}, 7:105--135, 2012.

\bibitem{Brock09}
A.~Brock, H.~Chang, and S.~Huang.
\newblock Non-genetic heterogeneity - a mutation-independent driving force for
  the somatic evolution of tumours.
\newblock {\em Nature Reviews - Genetics}, 10:336--342, 2009.

\bibitem{Burga12}
A.~Burga and B.~Lehner.
\newblock Beyond genotype to phenotype: why the phenotype of an individual
  cannot always be predicted from their genome sequence and the environment
  that they experience.
\newblock {\em FEBS Journal}, 279:3765--3775, 2012.

\bibitem{Codling08}
E.~A. Codling, M.~J. Plank, and S.~Benhamou.
\newblock Random walk models in biology.
\newblock {\em J. R. Soc. Interface}, 5:813--834, 2008.

\bibitem{Feinberg10}
A.~P. Feinberg and R.~A. Irizarry.
\newblock Stochastic epigenetic variation as a driving force of development,
  evolutionary adaptation, and disease.
\newblock {\em PNAS}, 107:1757--1764, 2010.

\bibitem{Gerlinger12}
M.~Gerlinger, A.~Rowan, S.~Horswell, J.~Larkin, D.~Endesfelder, E.~Gronroos,
  P.~Martinez, N.~Matthews, A.~Stewart, P.~Tarpey, I.~Varela, B.~Phillimore,
  S.~Begum, N.~Q. McDonald, A.~Butler, D.~Jones, K.~Raine, C.~Latimer, C.~R.
  Santos, M.~Nohadani, A.~C. Eklund, B.~Spencer-Dene, G.~Clark, L.~Pickering,
  G.~Stamp, M.~Gore, Z.~Szallasi, J.~Downward, P.~. Futreal, , and C.~Swanton.
\newblock Intratumor heterogeneity and branched evolution revealed by
  multiregion sequencing.
\newblock {\em The New England Journal of Medicine}, 366:883--892, 2012.

\bibitem{Giese03}
A.~Giese, R.~Bjerkvig, M.~Berens, and M.~Westphal.
\newblock Cost of migration: invasion of malignant gliomas and implications for
  treatment.
\newblock {\em J. Clin. Oncol.}, 21:1624--1636, 2003.

\bibitem{Greaves12}
M.~Greaves and C.~C. Maley.
\newblock Clonal evolution in cancer.
\newblock {\em Nature}, 481:306--313, 2012.

\bibitem{Guerrero12}
C.~Guerrero-Bosagna and M.~K. Skinner.
\newblock Environmentally induced epigenetic transgenerational inheritance of
  phenotype and disease.
\newblock {\em Molecular and Cellular Endocrinology}, 354:3--8, 2012.

\bibitem{Jeon10}
J.~Jeon, V.~Quaranta, and P.~T. Cummings.
\newblock An off-lattice hybrid discrete-continuum model of tumor growth and
  invasion.
\newblock {\em Biophysical Journal}, 98:37--47, 2010.

\bibitem{Kam09}
Y.~Kam, A.~Karperien, B.~Weidow, L.~Estrada, A.~R.~A. Anderson, and
  V.~Quaranta.
\newblock Nest expansion assay: a cancer systems biology approach to in vitro
  invasion measurements.
\newblock {\em BMC Res Notes}, 2(1):130--139, 2009.

\bibitem{Kam11}
Y.~Kam, K.~A. Rejniak, and A.~R.~A. Anderson.
\newblock Cellular modeling of cancer invasion: Integration of in silico and in
  vitro approaches.
\newblock {\em J Cell Physiol}, 227(2):431--438, 2011.

\bibitem{Kreso13}
A.~Kreso, C.~A. O'Brien, P.~{van Galen}, O.~I. Gan, F.~Notta, A.~M.~K. Brown,
  K.~Ng, J.~Ma, E.~Wienholds, C.~Dunant, A.~Pollett, S.~Gallinger,
  J.~McPherson, C.~G. Mullighan, D.~Shibata, and J.~E. Dick.
\newblock Variable clonal repopulation dynamics influence chemotherapy response
  in colorectal cancer.
\newblock {\em Science}, 339:543--548, 2013.

\bibitem{Lachmann96}
M.~Lachmann and E.~Jablonka.
\newblock The inheritance of phenotypes: an adaptation to fluctuating
  environments.
\newblock {\em J. Theor. Biol.}, 181:1--9, 1996.

\bibitem{Lee95}
Y.~Lee, S.~Kouvroukoglou, L.~V.\ McIntire, and K.~Zygourakis.
\newblock A cellular automaton model for the proliferation of migrating
  contact-inhibited cells.
\newblock {\em Biophysical Journal}, 69:1284--1298, 1995.

\bibitem{Ma11}
C.~Ma, R.~Fan, H.~Ahmad, Q.~Shi, B.~Comin-Anduix, T.~Chodon, R.~C. Koya, C.-C.
  Liu, G.~A. Kwong, C.~G. Radu, A.~Ribas, and J.~R. Heath.
\newblock A clinical microchip for evaluation of single immune cells reveals
  high functional heterogeneity in phenotypically similar t cells.
\newblock {\em Nature Medicine}, 17:738--744, 2011.

\bibitem{Mansury06}
Y.~Mansury, M.~Diggory, and T.~S. Deisboeck.
\newblock Evolutionary game theory in an agent-based brain tumor model:
  Exploring the 'genotype-phenotype' link.
\newblock {\em Journal of Theoretical Biology}, 238:146--156, 2006.

\bibitem{Marusyk12}
A.~Marusyk, V.~Almendro, and K.~Polyak.
\newblock {Intra-tumour heterogeneity: a looking glass for cancer?}
\newblock {\em Nature Reviews Cancer}, 12:323--334, 2012.

\bibitem{Marusyk13}
A.~Marusyk and K.~Polyak.
\newblock Cancer cell phenotypes in fifty shades of grey.
\newblock {\em Science}, 339:528--529, 2013.

\bibitem{Merlo10}
L.~M.~F. Merlo and C.~C. Maley.
\newblock The role of genetic diversity in cancer.
\newblock {\em J Clin Invest}, 120(2):401--403, 2010.

\bibitem{Niepel09}
M.~Niepel, S.~L. Spencer, and P.~K. Sorger.
\newblock Non-genetic cell-to-cell variability and the consequences for
  pharmacology.
\newblock {\em Current Opinion in Chemical Biology}, 13:556--561, 2009.

\bibitem{Quaranta09}
V.~Quaranta, D.~R. Tyson, S.~P. Garbett, B.~Weidow, M.~P. Harris, and
  W.~Georgescu.
\newblock Trait variability of cancer cells quantified by high-content
  automated microscopy of single cells.
\newblock {\em Methods Enzymol.}, 467:23--57, 2009.

\bibitem{Rando07}
O.~J. Rando and K.~J. Verstrepen.
\newblock Timescales of genetic and epigenetic inheritance.
\newblock {\em Cell}, 106(46):19352--19357, 2007.

\bibitem{Rowat09b}
A.~C. Rowat and D.~A. Weitz.
\newblock Understanding epigenetic regulation: Tracking protein levels across
  multiple generations of cells.
\newblock {\em European Physical Journal Special Topics}, 178:71--80, 2009.

\bibitem{Shannon48}
C.~E. Shannon.
\newblock A mathematical theory of communication.
\newblock {\em Bell Sys Tech J}, 27:379--423, 1948.

\bibitem{Simpson49}
E.~H. Simpson.
\newblock Measurement of diversity.
\newblock {\em Nature}, 163:688, 1949.

\bibitem{Sottoriva13}
A.~Sottoriva, I.~Spiteri, and S.~G.~M. Piccirillo.
\newblock Intratumor heterogeneity in human glioblastoma reflects cancer
  evolutionary dynamics.
\newblock {\em PNAS}, Early edition:1--6, 2013.

\bibitem{Sottoriva11}
A.~Sottoriva, L.~Vermeulen, and S.~Tavar\'{e}.
\newblock Modeling evolutionary dynamics of epigenetic mutations in
  hierarchically organized tumors.
\newblock {\em PLOS Computational Biology}, 7(5):1--11, 2011.

\bibitem{Spencer09}
S.~L. Spencer, S.~Gaudet, J.~G. Albeck, J.~M. Burke, and P.~K. Sorger.
\newblock Non-genetic origins of cell-to-cell variability in trail-induced
  apoptosis.
\newblock {\em Nature}, 459(7245):428--432, 2009.

\bibitem{Staudte97}
R.~G. Staudte, R.~M. Huggins, J.~Zhang, D.~E. Axelrod, and M.~Kimmel.
\newblock Estimating clonal heterogeneity and interexperiment variability with
  the bifurcating autoregressive model for cell lineage data.
\newblock {\em Mathematical Biosciences}, 143:103--121, 1997.

\bibitem{Swanton12}
C.~Swanton.
\newblock Intratumor heterogeneity: Evolution through space and time.
\newblock {\em Cancer Research}, 72(19):4875--4882, 2012.

\end{thebibliography}

\section*{Appendix 1: The heterogeneity index}
Common formulas for quantifying heterogeneity include those proposed by Simpson and Shannon in the late 1940's \cite{Shannon48,Simpson49}. Both of these values grow quickly as sample size increases and slower as the numbers even out. We want to avoid this bias and present a value that represents the occupied proportion of the available space. This point can be made clear with the figure below, which shows a distribution calculated by the heterogeneity index described here, which ranges from 0 to 1 linearly, and also by the normalized Shannon index, which is logarithmic, for comparison. 

\begin{figure}[ht!]
   \centering
    \includegraphics[width=\textwidth]{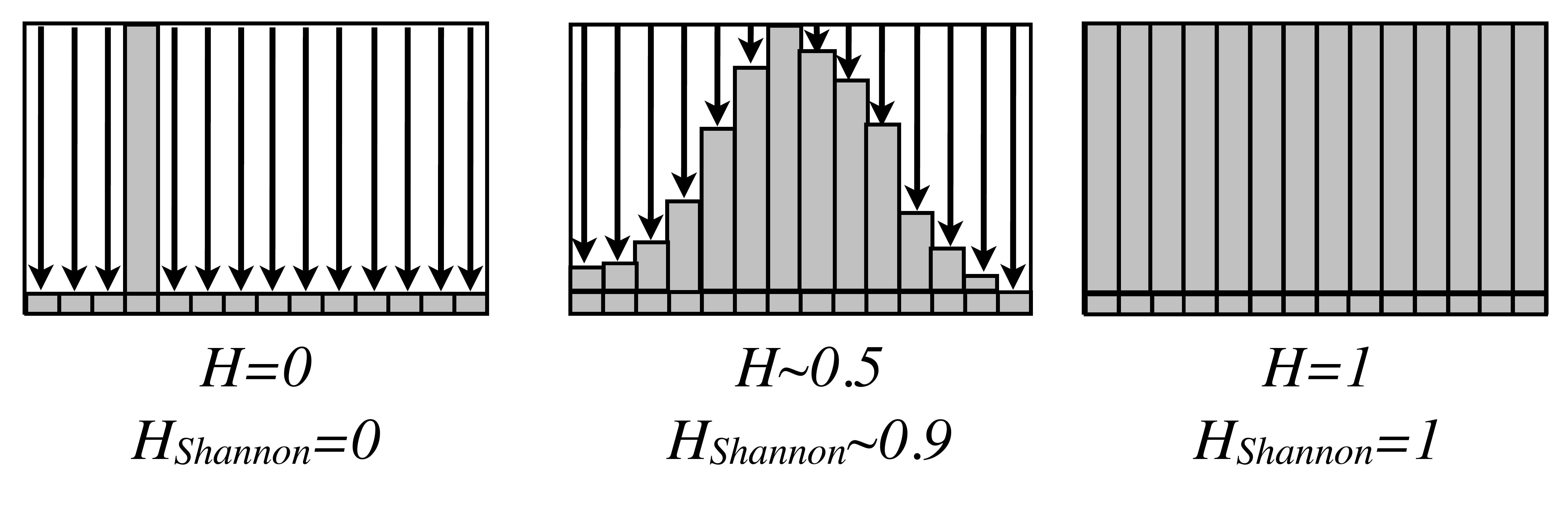} 
    \caption{An illustration of the heterogeneity index as calculated via Eq.\ \ref{hetIndexEq}, and with the Shannon index from the frequency distribution of trait values. If all cells in the population have the same trait value, we get a minimum heterogeneity index (left). If the population has an equal spread of trait values throughout the range, the heterogeneity index is maximum (right). The two values differ in the middle configuration, where the Shannon index grows logarithmically compared to the linear growth used here.}%
        \label{HETInd}
\end{figure}

Basically, we sum up over all bins, the difference between each bin and the bin with the highest frequency. So, if all traits occupy one bin, we get a maximum value, and if all bins are equally populated we get nil. We then subtract the summation from one to get the heterogeneity index. The equation for the heterogeneity index is as follows:

\begin{equation}
H=1-\frac{1}{N_{\rm max} (\tau_{\rm max}-\tau_{\rm min}-\Delta \tau)}\sum_{j} (N_{\rm max}-N_j) \Delta \tau
\label{hetIndexEq}
\end{equation}

This equation defines a simple measure for how disperse the values are within a trait's range $\tau \in (\tau_{\rm min},\tau_{\rm max})$. It ensures that $H=1$ when the range of trait values are filled equally and $H=0$ when all cells have the exact same trait value (c.f.\ Fig.\ \ref{HETInd}). We define $N_{\rm max}$ as the bin with the highest frequency, $N_j$ as the occupancy of every other bin, and $\Delta \tau$ as the bin size. Choosing the range and bin sizes appropriately is an important part of determining this value. 

\end{document}